\documentclass[a4paper,11pt]{article}
\pdfoutput=1 % if your are submitting a pdflatex (i.e. if you have
\usepackage{jcappub} % for details on the use of the package, please
\usepackage[T1]{fontenc} % if needed

\title{\boldmath Leptogenesis from spontaneous symmetry breaking during inflation}

\author[a]{Yi-Peng Wu,}
\author[b]{and Louis Yang,}
\author[b,c]{and Alexander Kusenko,}
\affiliation[a]{Research Center for the Early Universe (RESCEU),\\Graduate School of Science, The University of Tokyo,	Tokyo 113-0033, Japan}
\affiliation[b]{Kavli Institute for the Physics and Mathematics of the 
	universe (Kavli IPMU),\\WPI, UTIAS, The University of Tokyo, 5-1-5 Kashiwanoha, Kashiwa 277-8583, Japan}
\affiliation[c]{Department of Physics and Astronomy,	University of California, Los Angeles, 	\\California 90095-1547, USA}

\emailAdd{ypwu@resceu.s.u-tokyo.ac.jp}
\emailAdd{louis.yang@physics.ucla.edu}
\emailAdd{kusenko@ucla.edu}
\abstract{
Symmetry breaking in the Higgs field via a non-minimal coupling to gravity or higher-dimensional interactions with the inflaton can lead to condensation at a large vacuum expectation value (VEV) during inflation.  After inflation is over, the Higgs field must relax to the minimum of its effective potential, creating an era in which the CPT is effectively broken by the time-depended VEV.  We show that the matter-antimatter asymmetry can be generated during this relaxation epoch.  
}

\definecolor{linkcolor}{RGB}{41, 127, 255}
\hypersetup{
	pdfstartview={XYZ},
	colorlinks,
	allcolors=linkcolor,
	bookmarksopen
}

\begin{document}
	\maketitle
	\flushbottom

\section{Introduction}
%Spontaneous symmetry breaking plays a fundamental role in the applications of quantum field theory to nature. It provides a model-independent description for cosmic inflation, assuming that the observed primordial scalar fluctuation, $\zeta$, is sourced by the Goldstone mode released from the broken time-translation symmetry in the cosmic expansion \cite{Cheung:2007st,Senatore:2010wk,Noumi:2012vr} (see \cite{Baumann:2014nda,Piazza:2013coa} for recent reviews). The effective action in terms of the cosmological Goldstone mode includes a universal part completely fixed by the inflationary background. The universal action captures predictions from the whole class of single-field models with slow-roll assumptions, while the precision tests of the $\zeta$ power spectrum report no evidence beyond slow roll dynamics in the framework of single-field inflation \cite{Ade:2015lrj,Akrami:2018odb}.

The Higgs boson discovery has confirmed that the spontaneous symmetry breaking at the electroweak scale is generated by a scalar field~\cite{Aad:2015zhl}. The dynamics of this field at the end of inflation could be responsible for generation of the matter-antimatter asymmetry in the universe~\cite{Kusenko:2014lra,Pearce:2015nga,Yang:2015ida,Kusenko:2017kdr}.  This is because scalar fields develop large expectation values during inflation~\cite{Bunch:1978yq,Linde:1982uu,Affleck:1984fy,Starobinsky:1994bd}, and the subsequent relaxation of the Higgs field to the minimum of the effective potential can create a CP and CPT-violating background for the plasma, biasing the energy levels of particles and antiparticles, which leads to a successful leptogenesis~\cite{Kusenko:2014lra,Pearce:2015nga,Yang:2015ida,Kusenko:2017kdr}.  

In the case of the Higgs boson, the shape of the potential at large vacuum expectation value (VEV) is uncertain.  The reported central value of the Higgs mass $m_h \approx 125$ GeV \cite{Aad:2015zhl} suggests that the electroweak vacuum $v_{\rm EW} \simeq 246$ GeV could be a false vacuum during inflation \cite{EliasMiro:2011aa,Degrassi:2012ry,Buttazzo:2013uya,Hook:2014uia}, depending on the Hubble parameter during inflation, $H_I$, the top quark mass~\cite{Enqvist:2014bua}, the Higgs couplings with inflaton~\cite{Espinosa:2015qea,Herranen:2014cua}, the scheme of renormalization in the inflationary background~\cite{Kearney:2015vba,Herranen:2014cua,Espinosa:2015qea}, and the presence of higher-dimensional operators \cite{Branchina:2013jra} or new fields beyond the Standard Model \cite{EliasMiro:2012ay}.
However, as long as $H_I \gg v_{\rm EW}$, the high energy quantum fluctuations during inflation drive the Higgs field away from the Standard Model (SM) vacuum, while the spontaneous symmetry breaking asks $h$ to find $v_{\rm EW}$ after inflation ends. Regardless of the uncertainty from the underlying physics of inflation, the post-inflationary Higgs evolution towards the electroweak minimum can be an essential phase of the early Universe \cite{Enqvist:2013kaa,Kusenko:2014lra,Pearce:2015nga,Yang:2015ida,Figueroa:2015rqa}.

%Indeed, the Higgs evolution, relaxed from the value at the end of inflation, is an out-of-equilibrium process that satisfies all the necessary conditions for baryogenesis \cite{Sakharov:1967dj,Cohen:1987vi}.

The SM Higgs boson can be responsible for the observed matter-antimatter asymmetry, provided that (i) the Higgs field acquires a sufficiently large VEV, $h_0 \equiv \langle h \rangle$, and (ii) the mean variance $\langle h^2\rangle$ is small enough during inflation \cite{Kusenko:2014lra,Pearce:2015nga,Yang:2015ida,Kusenko:2017kdr}.
The first condition yields a suitable initial condition for the post-inflationary evolution of $h$ that can result in an effective chemical potential $\mu_{\rm eff}$ for the thermal equilibrium of the lepton number density. The chemical potential effectively shifts the ground state energy of lepton different from that of anti-lepton. With the assistance of the lepton number violating processes, a net lepton number can be generated,  provided that the chemical potential does not vanish before the reheating is completed. The second condition is to make sure that the final baryonic isocurvature perturbations due to the variance of $\mu_{\rm eff}$ at different patches of the Universe are compatible with the cosmic microwave background and big bang nucleosynthesis constraints~\cite{Kusenko:2017kdr,Inomata:2018htm}, given that $\mu_{\rm eff}$ is sourced by the time-derivative of $h$ with initial conditions depend on $\langle h^2\rangle$.

In this work, we consider a symmetry breaking in the Higgs potential led by the slow-roll dynamics of inflation in the framework of single-field models, where the inflaton sector is a scalar field, denoted by $\phi$, beyond the SM.
We show that the slow-roll dynamics of inflation can trigger a redistribution of the Higgs condensate that satisfies both  conditions (i) and (ii) for a successful {\it relaxation leptogenesis} \cite{Kusenko:2014lra,Pearce:2015nga,Yang:2015ida,Kusenko:2017kdr}). 

Examples of symmetry breaking in the SM sector during or immediately after inflation are widely discussed in the literature~\cite{He:2018gyf,Kumar:2017ecc,Wu:2018lmx,Chen:2018xck,Graham:2015cka,He:2018mgb,Opferkuch:2019zbd,GarciaBellido:2001cb}. One of the interesting consequences of the symmetry breaking is that the gauge fields in general can obtain masses of $\mathcal{O}(H_I)$, which results in the "heavy-lifting" of the SM mass spectrum \cite{Kumar:2017ecc,Wu:2018lmx}. In this work we focus on a class of symmetry breaking phenomena produced by the Higgs coupling with the kinetic term of the inflaton, where the Higgs-inflaton coupling is a higher-dimensional operator suppressed by some cutoff scale $\Lambda$. This non-canonical type of kinetic interactions can introduce an effective tachyonic mass in the Higgs potential and generate some non-trivial $h_0$ and $m_h$ associated with the normalized inflaton velocity $\dot{\phi}/\Lambda$ \cite{Kumar:2017ecc,Wu:2018lmx}. Given that the amplitude of the $\zeta$ power spectrum, $P_\zeta = H_I^4/(2 \dot{\phi}^2)$, is fixed by observations, the Higgs VEV (or mass) is therefore fixed by the scales of $H_I$ and $\Lambda$. 

When the inflaton makes a transition from the slow-roll dynamics to rapid oscillations for (p)reheating, the sudden increase in the kinetic energy density of the inflaton leads to a significant enhancement of the tachyonic mass in the Higgs potential. Initial conditions from enhanced tachyonic instability lead to a novel relaxation dynamics for the post-inflationary Higgs evolution. We will discuss  the Higgs relaxation treating the effects of the inflaton as an external driving force. The additional energy form the inflaton sector is transferred to the Higgs boson, and the amplitude of the Higgs oscillations can be much greater than the initial VEV $h_0$. This is in contrast with a free relaxation as previously considered in Refs.~\cite{Kusenko:2014lra,Yang:2015ida}, where the initial value $h_0$ was the maximal amplitude of the oscillation. We show that the enhanced amplitude in the forced relaxation process enlarges the parameter space for a successful leptogenesis significantly. 

The paper is organized as the follows. In Section~\ref{Sec. sym_breaking}, we briefly review the single-field inflation with a symmetry breaking motivated by the heavy-lifting mechanism. In Section~\ref{Sec. Higgs_relax}, we discuss the Higgs evolution during reheating and consider one example of a ``free-fall relaxation'' and an example of the ``forced relaxation.'' After clarifying our set up for the reheating process, we apply the initial conditions from the forced relaxation and describe a  successful leptogenesis in Section~\ref{Sec. leptogenesis}. Finally, we give our conclusions in Section~\ref{Sec. conclusion}.

\section{Inflation with symmetry breaking}\label{Sec. sym_breaking}
In this section we review inflationary scenarios associated with a gauge symmetry breaking in the SM sector.  The Higgs boson  acquires a non-trivial vacuum expectation value (VEV), $h_0 \equiv \langle h\rangle \neq 0$, where $h$ is the neutral component of the Higgs doublet $\Phi_H = (0, h)^T/\sqrt{2}$.  We focus on non-Higgs inflation scenarios, and we consider the inflaton, $\phi$, in the framework of single-field inflation  with slow-roll conditions. If the symmetry breaking is triggered by the slow-roll dynamics of inflation, the non-zero VEV $h_0$ then characterizes a modified Higgs scale associated with the inflationary Hubble parameter $H_I$. As a result of symmetry breaking, SM particles gain masses of $\mathcal{O}(H_I)$ and behave as heavy degrees of freedom during inflation.
This is sometimes called the heavy-lifting of the SM mass spectrum \cite{Kumar:2017ecc,Wu:2018lmx}.

One of the simplest realizations for the symmetry breaking during inflation is to introduce a tachyonic mass to the Higgs sector. Following the principles of the effective field theory (EFT) expansion, theories for the heavy-lifting can be summarized as
\begin{align}\label{EFT}
\mathcal{L} = \sqrt{-g} \left[ \xi R \Phi^\dagger_H\Phi_H - \lambda(\Phi_H^\dagger \Phi_H)^2 + \sum_i \frac{c_i}{\Lambda^{d_i -4}} \mathcal{O}_i\right],
\end{align}
where $\Lambda$ is the cutoff for higher-order interactions, $d_i$ is the mass dimensions for the effective operators $\mathcal{O}_i$ made from Higgs and inflaton, and $c_i$ are $\mathcal{O}(1)$ parameters.  Assuming an approximated shift symmetry in the inflaton sector during the slow roll, $\mathcal{O}_i$ only contain derivative couplings of $\phi$. For simplicity, we approximate $\lambda \lesssim 10^{-2} $ by a positive constant.

The first choice for a tachyonic mass is a ``wrong-sign'' non-minimal coupling $\xi$. Since $R \approx 12 H_I^2$ during inflation, the non-minimal coupling gives rise to a symmetry breaking with an effective Higgs mass $m_h \sim \sqrt{12 \xi} H_I/\sqrt{\lambda}$. Removing the non-minimal coupling by a conformal transformation into the Einstein frame, one can see that the tachyonic mass is effectively given by the inflaton potential \cite{Kumar:2017ecc}. In this work we shall focus on the second possibility where the symmetry breaking is induced by the Higgs-inflaton coupling $\mathcal{O}_i$. To simplify the discussion, we set $\xi =0$, although this condition is not necessary in a more general setup. A stable VEV $h_0$ develops for the dimension 6 operator of the form~\cite{Wu:2018lmx,Chen:2018xck,Kumar:2017ecc}
\begin{align}\label{toy model}
\mathcal{O}_{\phi h} = - \frac{(\partial\phi)^2}{\Lambda^2} \Phi^\dagger_H\Phi_H = - \frac{h^2}{2\Lambda^2} (\partial\phi)^2,
\end{align}
which respects the shift symmetry for $\phi$.
In the homogeneous background, $(\partial\phi)^2 = -\dot{\phi}_0^2$ acts as a tachyonic mass for $h$, and $\mathcal{O}_{\phi h}$ leads to a symmetry breaking with $h_0 = \pm \dot{\phi}_0/(\sqrt{\lambda}\Lambda)$ so that the effective mass $m_h = \sqrt{2}|\dot{\phi}_0|/\Lambda$. 

To ensure a well-defined EFT expansion, we require that $h_0^2 \ll \Lambda^2$, which also guarantees $\vert\dot{\phi}_0 \vert\ll \Lambda^2$ as $\lambda\ll 1$. The condition  $h_0^2/\Lambda^2 < 1$ specifies the parametric space of \textit{decoupling}, where the Higgs corrections to the slow-roll inflation are negligible. The power spectrum in this regime is thus given by the standard single-field inflation \cite{Wu:2018lmx}:
\begin{align}\label{power_spectrum}
P_\zeta = 2\pi^2 A_s = \frac{H_I^4}{2\Lambda^2 \dot{\theta}^2_0},
\end{align}
where $A_s = 2.2 \times 10^{-9}$ is the spectrum amplitude of the curvature perturbation $\zeta$ and we have defined the dimensionless parameter $\theta \equiv \phi/\Lambda$ for convenience.

\section{Post-inflationary Higgs relaxation}\label{Sec. Higgs_relax}
The slow-roll inflation is interrupted when the inflaton rolls into a deep potential valley, and rapid oscillations ensue. 
In this section we identify two classes of post-inflationary Higgs dynamics triggered by such a phase transition. In the first class of scenarios, the tachyonic mass that builds a barrier around the origin of the Higgs potential is diluted by the cosmic expansion and is eventually eliminated by the finite temperature effects. The homogeneous Higgs condensate $h_0$ then starts to move toward the potential minimum as its vacuum energy overcomes the potential barrier. We refer to the Higgs evolution in these scenarios as the ``free-fall relaxation.'' Leptogenesis from the free-falling Higgs relaxation has been investigated in Refs.~\cite{Kusenko:2014lra,Yang:2015ida}.  In the second class of scenarios, the potential barrier (built up by the tachyonic mass term) can be enhanced by the transition of the inflaton dynamics from a slow roll to rapid oscillation. The Higgs condensate is forced to roll away from the potential origin at the beginning of the post-inflationary epoch (reheating). In such a scenario the Higgs field receives some additional energy from inflaton.  This kind of the Higgs evolution is referred to as ``forced relaxation.'' Leptogenesis from forced relaxation has not been discussed in the literature.

\begin{figure}
	\begin{center}
		\includegraphics[width=50mm]{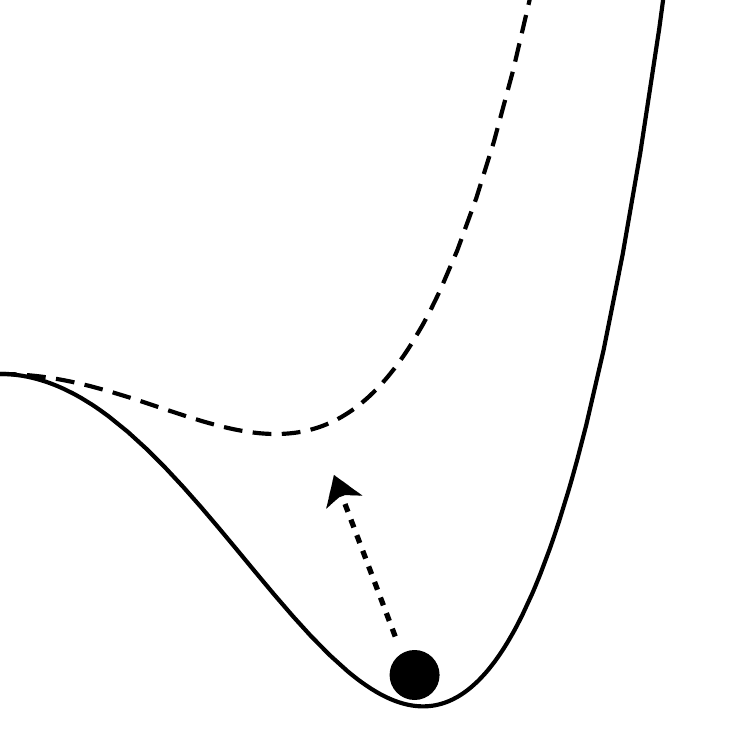}
		\qquad\qquad\qquad
		\includegraphics[width=51.5mm]{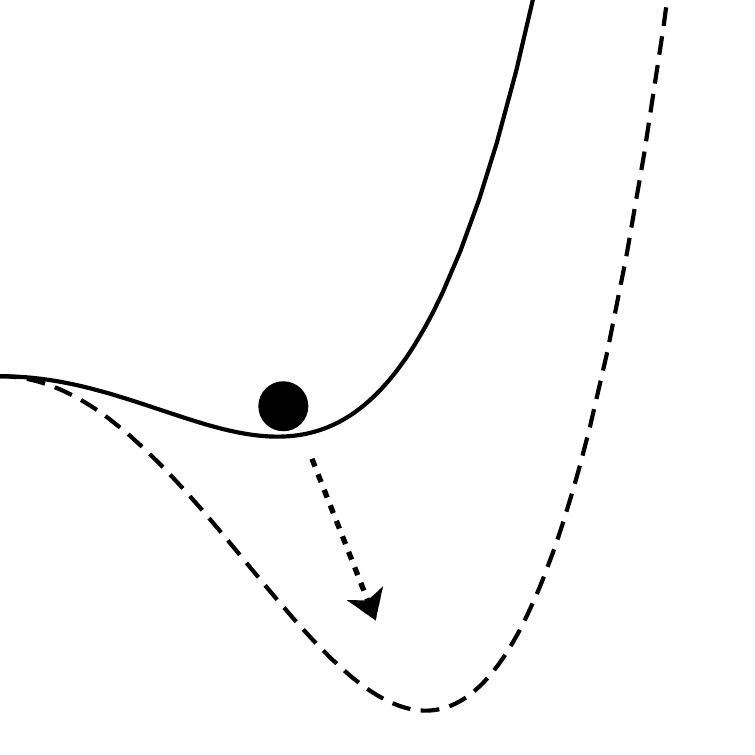}
	\end{center}
	\caption{\label{fig_Higgs_potential}Illustration of the effective Higgs potential from the end of inflation (solid-curve) to preheating (dashed-curve). Left panel: the Higgs field starts to roll down towards the origin due to the decrease of the barrier (free-fall relaxation).  Right panel: the Higgs field starts to roll away from the origin due to the enhancing barrier from the tachyonic mass term (forced relaxation).
	}
\end{figure}

\subsection{Free-fall relaxation}

After inflation, the Higgs condensation at a non-zero value signals that the system is not in equilibrium. In the absence of external forces, the Higgs field simply relaxes to the minimum at the origin. Initial conditions for this class of relaxation may come from the accumulation of long-wavelength fluctuations in a shallow potential with only the quartic self-coupling $\lambda h^4/4$, where the mean Higgs VEV $\sqrt{\langle h^2\rangle} \sim H_I/\lambda^{1/4}$ is developed even without symmetry breaking \cite{Starobinsky:1994bd}. The Higgs condensate relaxes from this initial VEV in the post-inflationary epoch when the Hubble friction is reduced. One can also consider a relaxation from the false vacuum driven by a negative running values of the self-coupling $\lambda < 0$.  This makes the potential unbounded from below, but this problem can be rectified by some higher-dimensional operators. In this case, the Higgs motion is triggered by the finite-temperature effects during reheating. See~\cite{Kusenko:2014lra,Yang:2015ida} for a detailed study of these two types of initial conditions.

To illustrate explicitly the Higgs relaxation in the free-fall class, we consider as an example where the symmetry breaking of the Higgs potential is led by a wrong-sign non-minimal coupling $\xi R h^2/2$ in \eqref{EFT}. Omitting $\phi$-$h$ interactions and higher-order effective operators, the homogeneous field equations are given by \cite{Geng:2015nnb}:
\begin{align} 
\label{eq:Friedmann_free1}
3M_p^2H^2 &= \frac{1}{2}  \dot{\phi}^2 +\frac{1}{2}\dot{h}^2 +V_{\rm osc}(\phi) + \frac{\lambda}{4} h^4 -3\xi H^2 h^2 - 6\xi H \dot{h}h, \\
\label{eq:Friedmann_free2}
-2M_p^2 \dot{H} &=   \dot{\phi}^2 + (1 + 2\xi)\dot{h}^2 + 2\xi \left(\ddot{h} + \dot{H}h - H\dot{h}\right) h, 
\end{align}
where $V_{\rm osc}(\phi)$ describes the inflaton potential for rapid oscillations. At the beginning of the post-inflationary epoch, the Hubble rate is very large so that one can neglect the perturbative decay of the inflaton. The evolution of the homogeneous parts of $\phi$ and $h$ are governed by the equations 
\begin{align} \label{eom_free:theta}
\ddot{\phi} +3 H\dot{\phi} +  V_\phi&= 0, \\
\ddot{h} + 3H \dot{h} + \lambda h^3 -6\xi \left(\dot{H} + 2H^2\right) h &= 0  , \label{eom_free:h}
\end{align} 
where $V_\phi \equiv \partial V_{\rm osc}/ \partial\phi$. The initial value of the Higgs VEV is $h_0 = \sqrt{12\xi} H_I/\sqrt{\lambda}$, which can be obtained from the effective potential during inflation $V_{\rm eff} (h) = \lambda h^4/4 - 6\xi H_I^2 h^2$ where $R = 12 H_I^2$. We assume both $\xi$ and $\lambda$ are positive constants for simplicity.

\begin{figure}
	\begin{center}
		\includegraphics[width=12cm]{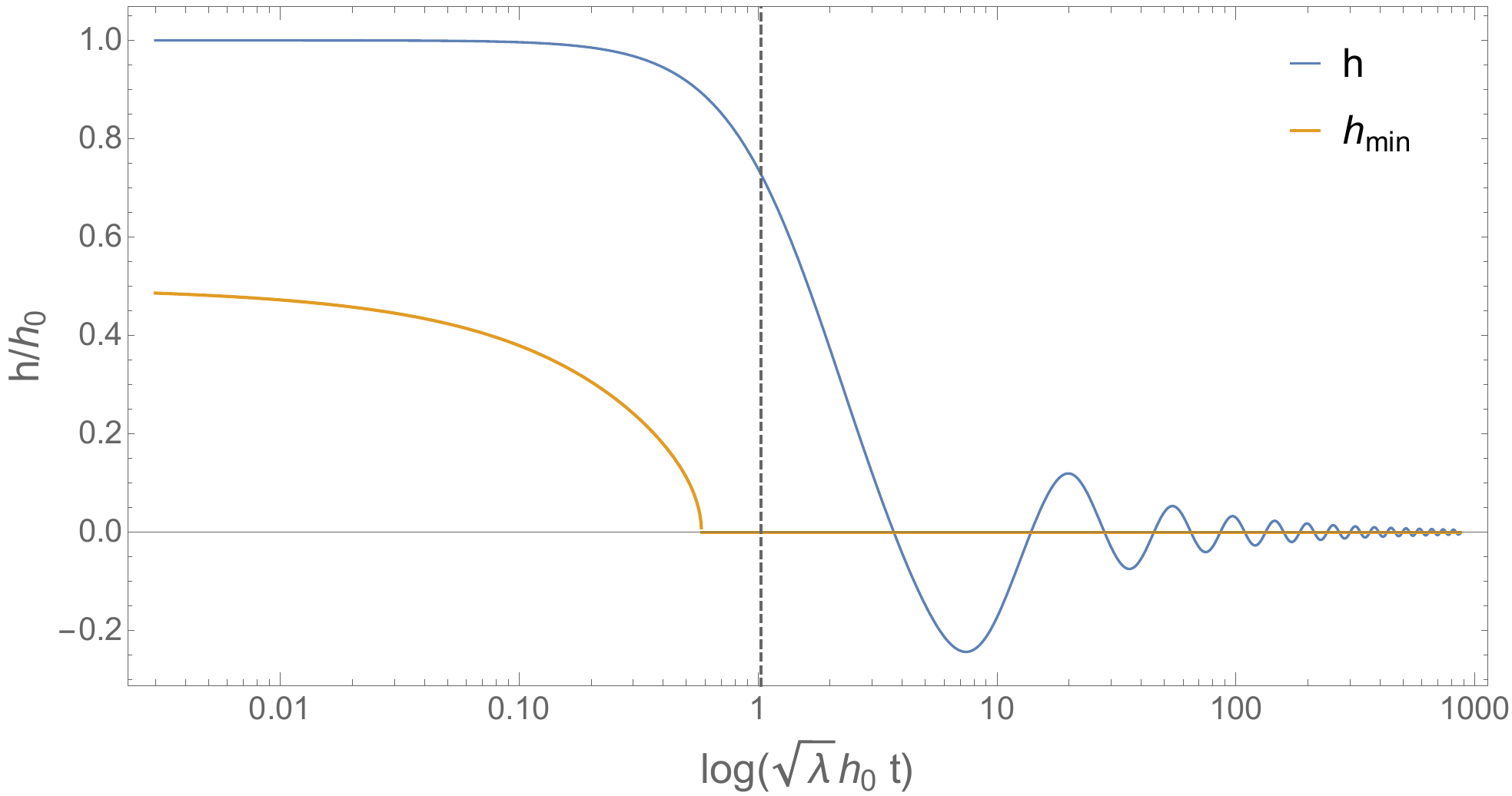} 
		%\par
	\end{center}
	\caption{Higgs relaxation from free-fall initial conditions. The parameters are $\Lambda_I=10^{16}\,\text{GeV}$, $\Gamma_{I}=10^{6}\,\text{GeV}$, and $\lambda=0.001$. The blue line shows the evolution of the Higgs VEV during reheating. The yellow line shows the positive minimum of the Higgs potential. The dashed line indicates the time when maximum temperature is reached. \label{fig:nonmini_Higgs-evolution}}
\end{figure}

It is convenient to approximate the inflaton dynamics by the standard harmonic oscillations, so that the averaged density behaves as that of pressureless matter. 
%This assumption is ture when the all the terms involved with Higgs are always subdominant densities. 
We shall choose the time scale of the inflaton oscillations to be very different from that of the Higgs oscillation such that the Higgs dynamics can be decoupled from the background evolution. We assume that reheating proceeds via the perturbative decay of the inflaton into radiation with a decay width $\Gamma_{I}$. During reheating, the tachyonic mass term due to the non-minimal coupling is gradually cancelled out by the finite-temperature effect, as illustrated in the left-panel of Figure~\ref{fig_Higgs_potential}. To show the evolution of the Higgs VEV, we provide a numerical example in Figure \ref{fig:nonmini_Higgs-evolution} with the energy scale of inflation $\Lambda_{I} = (3M_p^2 H_I^2)^{1/4} = 10^{16}$ GeV. We take $\Gamma_{I} = 10^6$ GeV and the background evolution is described by equations 
\begin{align}
\dot{\rho}_\phi + 3H \rho_\phi + \Gamma_{I}\rho_\phi = 0, \qquad \dot{\rho}_R + 4H \rho_{R} = \Gamma_{I} \rho_\phi,
\end{align}
where $3M_p^2 H^2 \simeq \rho_\phi + \rho_{R}$ and $\rho_{R}$ is the density of radiation. We have included in Figure~\ref{fig:nonmini_Higgs-evolution} the finite-temperature effects with detail explanation given in the next section. One can see that the barrier produced by the non-minimal coupling is eliminated by the finite-temperature effects around the time of the maximum reheating $t_{\rm max} \sim 2/(3H_I)$ \cite{Kusenko:2017kdr}, at which point the Higgs condensate starts to move from its initial condition set by inflation.

\subsection{Forced relaxation}
Let us now discuss the Higgs dynamics spanned by the $\phi$-$h$ interactions.  We will treat the inflaton as an external driving force during the relaxation of the Higgs VEV.  We consider a symmetry breaking led by the operator \eqref{toy model} so that both the Higgs VEV and the Higgs mass rely on the inflaton velocity. 
%In this section we address the dynamic of the $\phi$-$h$ system after inflation ends.
With the presence of the higher-order coupling $\mathcal{O}_{\phi h}$, the background field equations for the two-field system are  
\begin{align} 
\label{eq:Friedmann_1}
3M_p^2H^2 &= \frac{1}{2} \left(1+ \frac{h^2}{\Lambda^2}\right) \dot{\phi}^2 +\frac{1}{2}\dot{h}^2 +V_{\rm osc}(\phi) + \frac{\lambda}{4} h^4, \\
\label{eq:Friedmann_2}
-2M_p^2 \dot{H} &= \left(1+ \frac{h^2}{\Lambda^2}\right)  \dot{\phi}^2 + \dot{h}^2, 
\end{align}
where $V_{\rm osc}(\phi)$ describes the inflaton potential in the post slow-roll epoch. We take $\xi =0$ for simplicity. The equations of motion for $\phi$ and $h$ are
\begin{align} \label{eom:theta}
\ddot{\phi} +3 H\dot{\phi} + \frac{2h\dot{h}}{\Lambda^2+ h^2} \dot{\phi}+   \frac{V_\phi}{\Lambda^2+ h^2}&= 0, \\
\ddot{h} + 3H \dot{h} + \lambda h^3 &=  h \frac{\dot{\phi}^2}{\Lambda^2}, \label{eom:h}
\end{align} 
where $V_\phi \equiv \partial V_{\rm osc}/ \partial\phi$. Let us  introduce the energy scale of inflation $\Lambda_{I}^4 = 3 M_p^2 H_I^2$. One can solve the system of equations with the initial condition $V_{\rm osc}(\phi_0) = \Lambda_{I}^4$ for a suitable value of $\phi_0$, given that the Higgs density $\rho_h = \lambda h_0^4/4$, and the kinetic terms in \eqref{eq:Friedmann_1} are negligible during inflation. From the equation of motion \eqref{eom:h} one can deduce that the effective Higgs potential in this scenario is 
\begin{align}
V_{\rm eff}(h) = \frac{1}{4} \lambda h^4 - \frac{1}{2}\dot{\theta}^2 h^2,
\end{align}
where, using $\dot{\theta} = \dot{\phi}/\Lambda$. The inflaton velocity acts as a centrifugal force and leads to the symmetry breaking.\footnote{By the field redefinitions $\theta \equiv \phi/\Lambda$ and $R\equiv (\Lambda^2 + h^2)^{1/2}$, the kinetic terms of the inflaton can be canonically normalized in the polar representation. In this representation, the symmetry breaking in the radial mode is manifest as $h^4 = (R^2 -\Lambda^2)^2$; see more discussion in \cite{Wu:2018lmx}.}
 Since energy conservation requires that the expectation value $\dot{\theta}_{\rm osc}$ during reheating be much greater than $\dot{\theta}_0$, the tachyonic mass in the Higgs potential  increases and triggers the evolution of the Higgs field. The change of the Higgs potential from inflation to reheating for this scenario is illustrated in the right-panel of Figure~\ref{fig_Higgs_potential}.

\subsubsection{Preheating}

%Without a loss of generality, 
As a simple possibility, we consider that inflaton decay into radiation through a perturbative channel with a decay width $\Gamma_{I}$. 
\footnote{Reheating could occur via a non-perturbative channel due to resonance between Higgs and inflaton, see also Appendix \ref{Appendix}. However, at most of the time the Higgs field during forced relaxation is very massive except for the periods around crossing through the origin. A careful study for the Higgs production rate in the scenario of forced relaxation is required as a future effort.}
At the beginning of the post-inflation era, $H \gg \Gamma_{I}$ so that we can neglect the effect of the inflaton decay.
 We approximate near the bottom of the valley of the potential as $V_{\rm osc}(\phi) = \frac{1}{2}m_\phi^2 \phi^2$, where $m_\phi$ is the inflaton mass during (p)reheating. We require $m_\phi \gg H_I$ where $H_I$ is the Hubble scale of  inflation. It is interesting to investigate the joined Higgs-inflaton evolution in the decoupling limit ($h_0 \ll \Lambda$) or in the large-mixing limit ($h_0 \gtrsim \Lambda$).
{\footnote{To justify the investigation in the large-mixing limit with $h_0/\Lambda > 1$, one would need to assume a fine cancellation of higher-order terms in the EFT expansion \cite{Wu:2018lmx}. It is interesting to consider a practical model to realize this setup.}}

\begin{figure}
	\begin{center}
		\includegraphics[width=70mm]{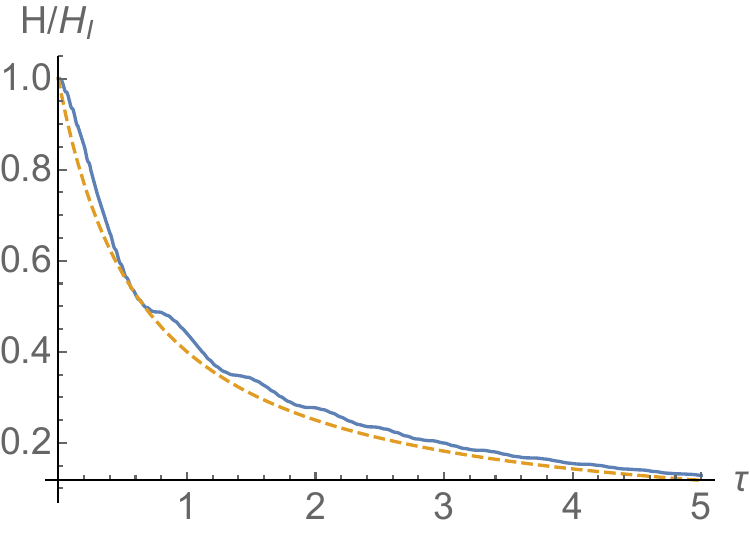}
		\hfill
		\includegraphics[width=71mm]{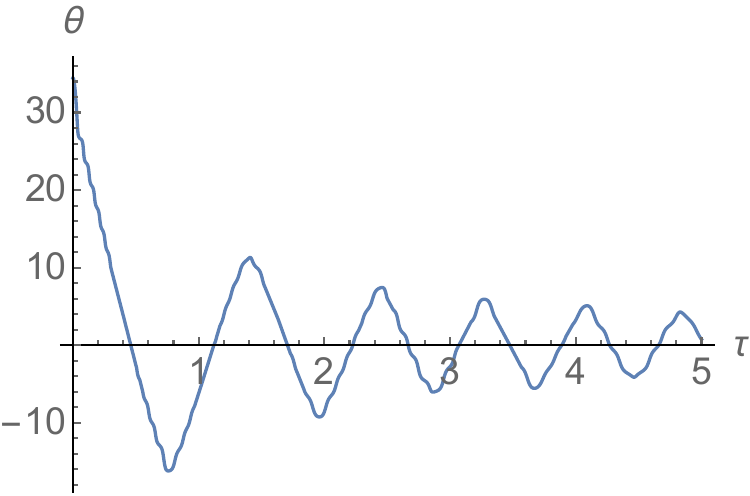}
	\end{center}
	\caption{\label{fig_weak-coupling}The evolution of the Hubble parameter (left panel) and the inflaton value (right panel) with respect to $\tau = H_I \,t$ in the case with $h_0 / \Lambda = 0.08$. The dashed line in the left panel is the Hubble evolution of the standard matter domination. 
	}
\end{figure}

\begin{figure}
	\begin{center}
		%\includegraphics[width=50mm]{higgs_potential.pdf}
		%\hfill
		\includegraphics[width=70mm]{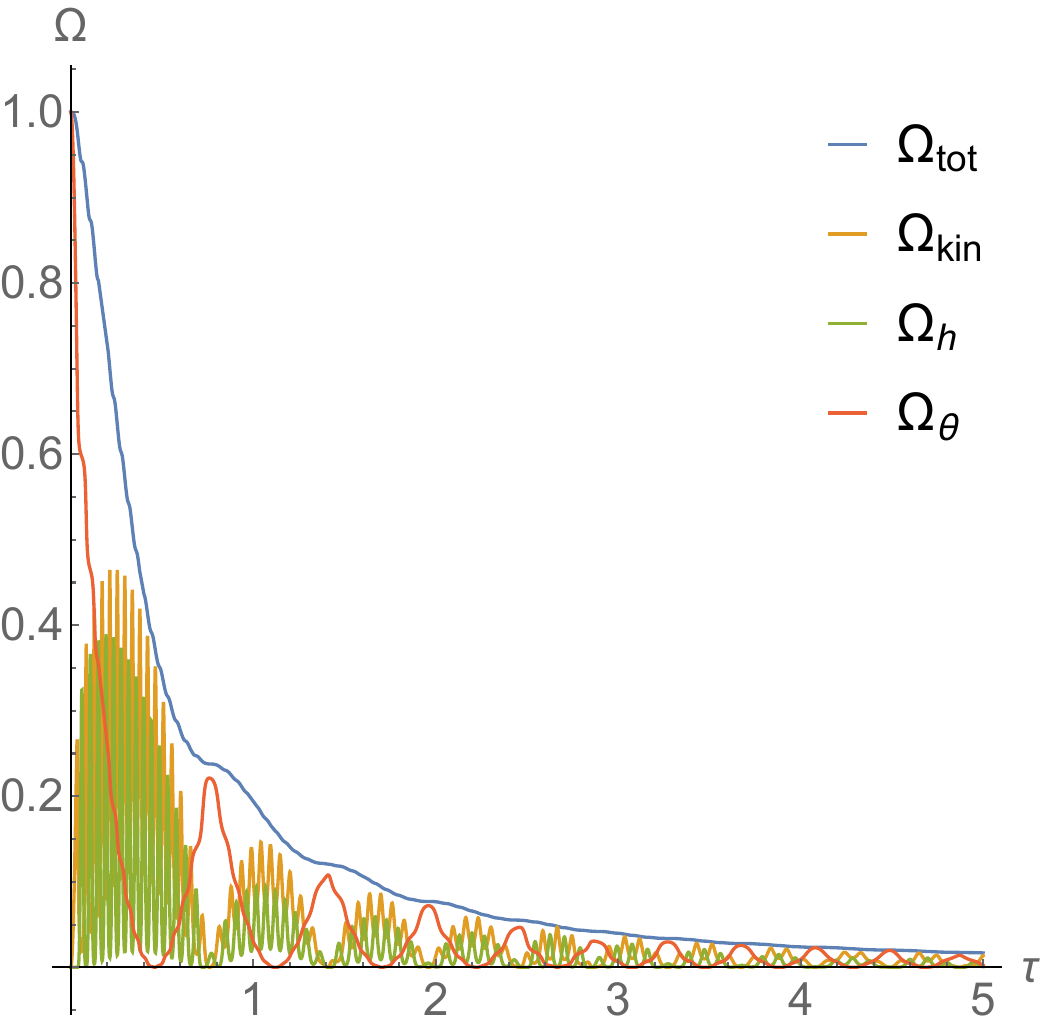}
		\hfill
		\includegraphics[width=70mm]{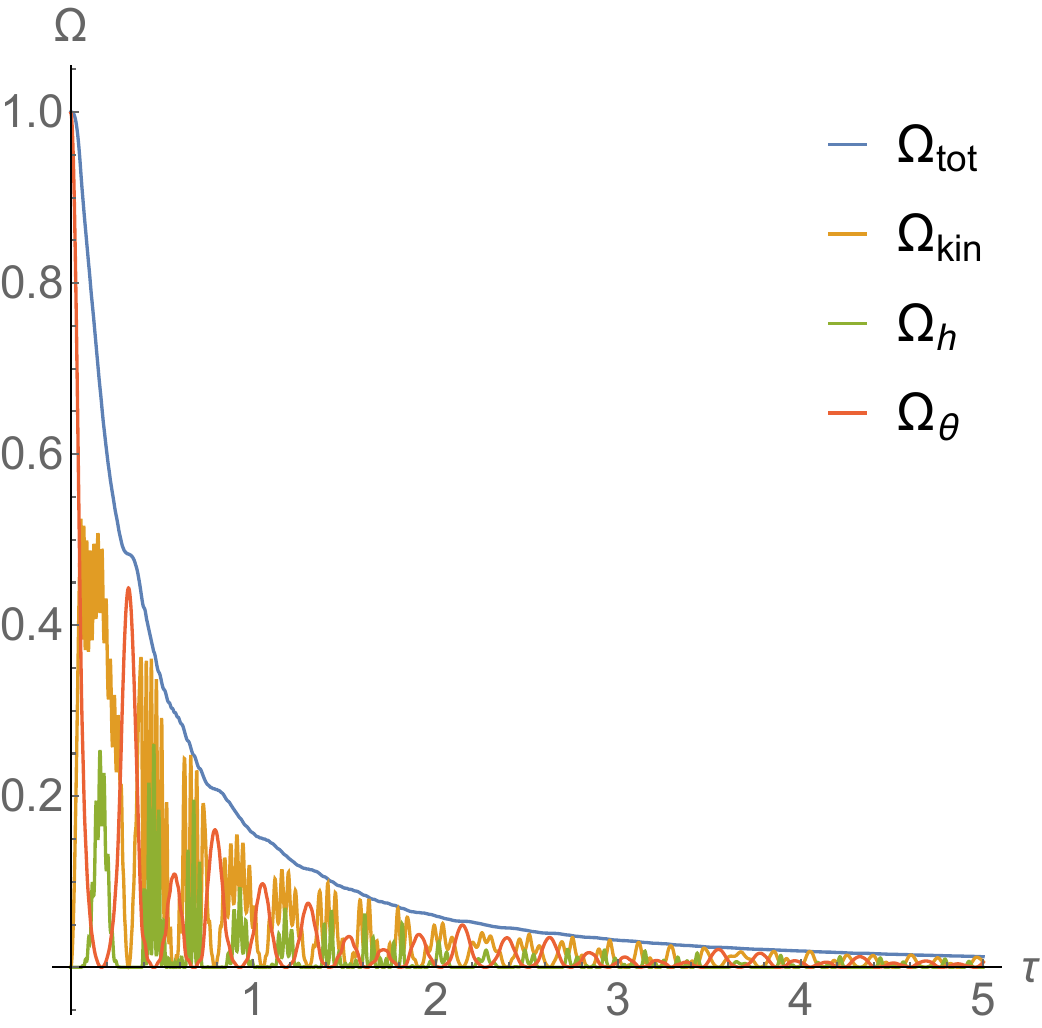}
		%\hfill
		%\includegraphics[width=50mm]{density_tau_sc2.pdf}
	\end{center}
	\caption{\label{fig_density_wc}The evolution of the density fraction $\Omega_i = \rho_i /\Lambda_I^4$ with respect to $\tau = H_I \,t$ in the decoupling case with $h_0 / \Lambda = 0.08$. Left-panel: the evolution based on \eqref{eom:theta} and \eqref{eom:h}. Right-panel: the evolution based on \eqref{eom:shift_symm_breaking} and \eqref{eom:shift_symm_breaking_h} with the presence of a portal coupling $g_s h^2\phi^2/2$.
	}
\end{figure}

\bigskip\noindent
\textbf{Decoupling.}
In the case of $h_0 \ll \Lambda$, the Higgs contribution to the background dynamics of the inflaton is negligible. Keeping the lowest order in $h_0/\Lambda$, the equation of motion \eqref{eom:theta} ca be reduced to 
\begin{align}
\ddot{\theta} +3 H\dot{\theta} +  V_\theta/\Lambda^2 = 0,
\end{align}
which is nothing but the equation of the standard single-field inflation. Thus the oscillations in a quadratic potential $V_{\rm osc}$ are expected to produce a background that scales as matter, as shown in Fig. \ref{fig_weak-coupling}.
Nevertheless, the inflaton oscillations are  not perfectly harmonic since some energy is taken away by Higgs when the total energy converts to the kinetic component $\rho_{\rm kin} = (\Lambda^2 + h^2)\dot{\theta}^2/2$ in \eqref{eq:Friedmann_1}. As shown in Fig. \ref{fig_density_wc}, the Higgs density can take an important fraction of the total at the moment when $\rho_{\rm kin}$ dominates. The dimensionless density fraction is defined as $\Omega_i = \rho_i /\Lambda_I^4$. The oscillating time scale of Higgs is characterized by the effective mass $m_h \sim \sqrt{2}\dot{\theta}_{\rm osc}$ where $\dot{\theta}_{\rm osc}(t_0)\approx \Lambda_I^2/\Lambda$. Figures \ref{fig_weak-coupling} and \ref{fig_density_wc} exhibit an example with $m_h \gg m_\phi$, where the oscillating time scale of the inflaton is much longer than that of Higgs.
Note that the potential energy is defined as $\rho_\theta = V_{\rm osc}(\phi) = m_\phi^2\Lambda^2 \theta^2/2$, and the total density $\Omega_{\rm tot}$ is $3M_p^2H^2/\Lambda_I^4$.

\begin{figure}
	\begin{center}
		\includegraphics[width=70mm]{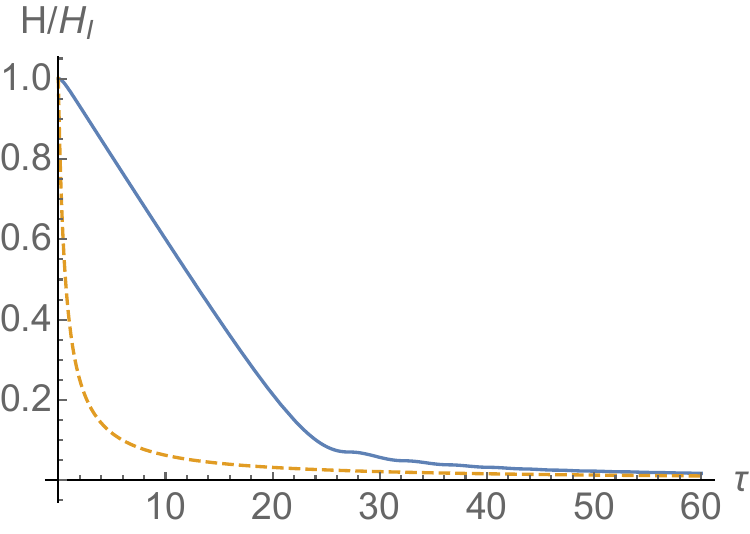}
		\hfill
		\includegraphics[width=71mm]{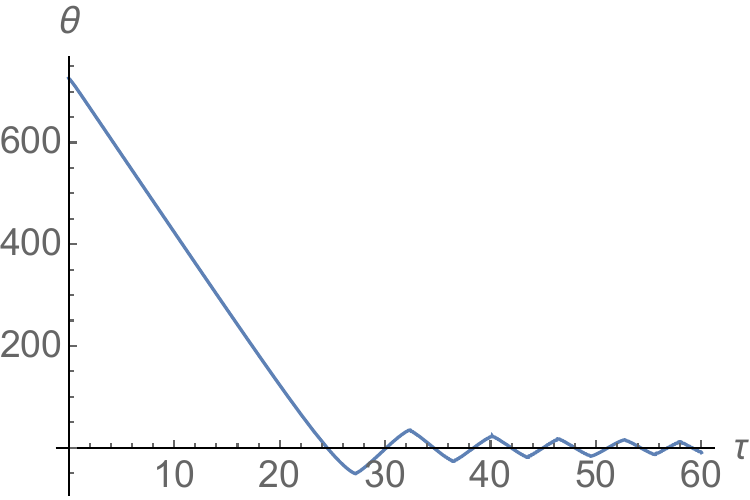}
	\end{center}
	\caption{\label{fig_strong-coupling}The evolution of the Hubble parameter (left panel) and the inflaton value (right panel) with respect to $\tau = H_I \,t$ in the case with $h_0 / \Lambda = 67$. The dashed line in the left panel is the Hubble evolution of the standard matter domination. 
	}
\end{figure}

\begin{figure}
	\begin{center}
		%\includegraphics[width=50mm]{higgs_potential.pdf}
		%\hfill
		%\includegraphics[width=50mm]{density_tau_dc.pdf}
		%\hfill
		\includegraphics[width=70mm]{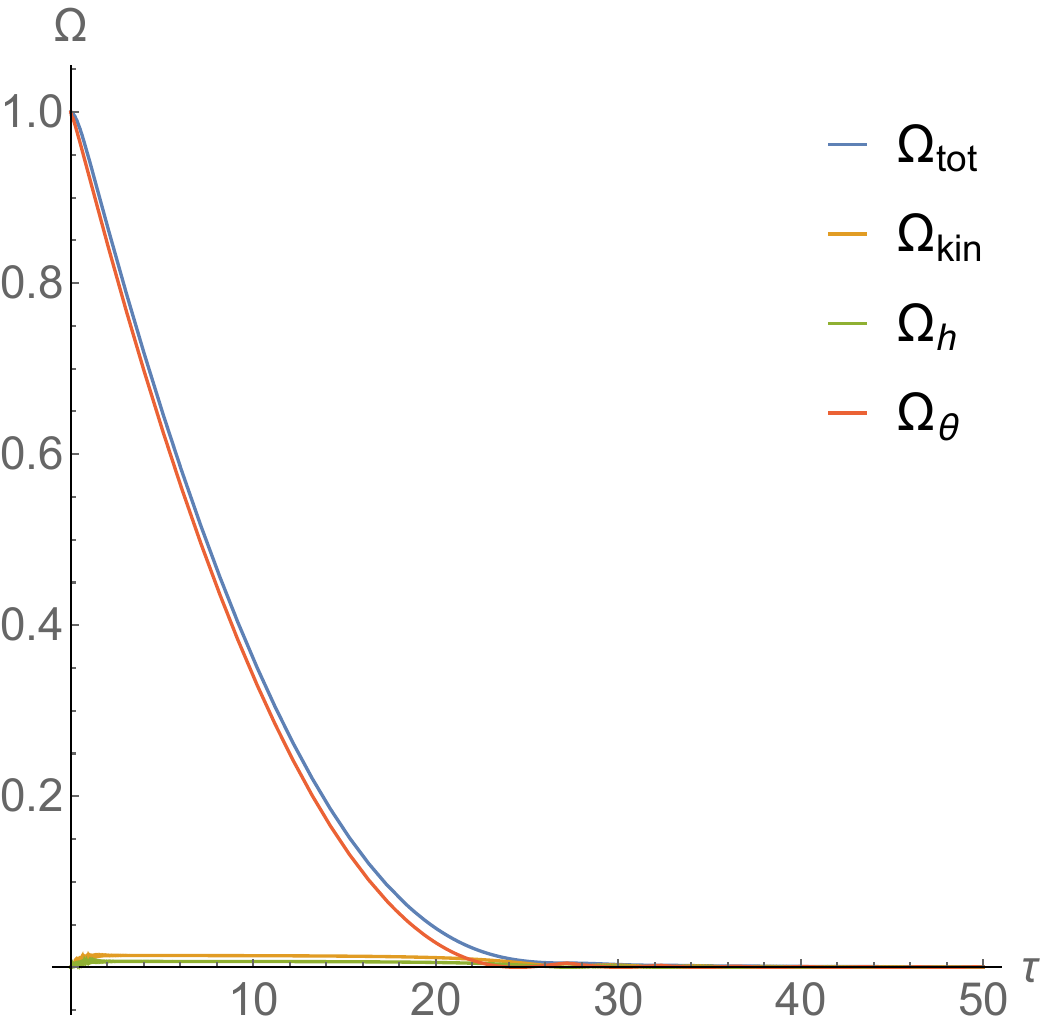}
		\hfill
		\includegraphics[width=70mm]{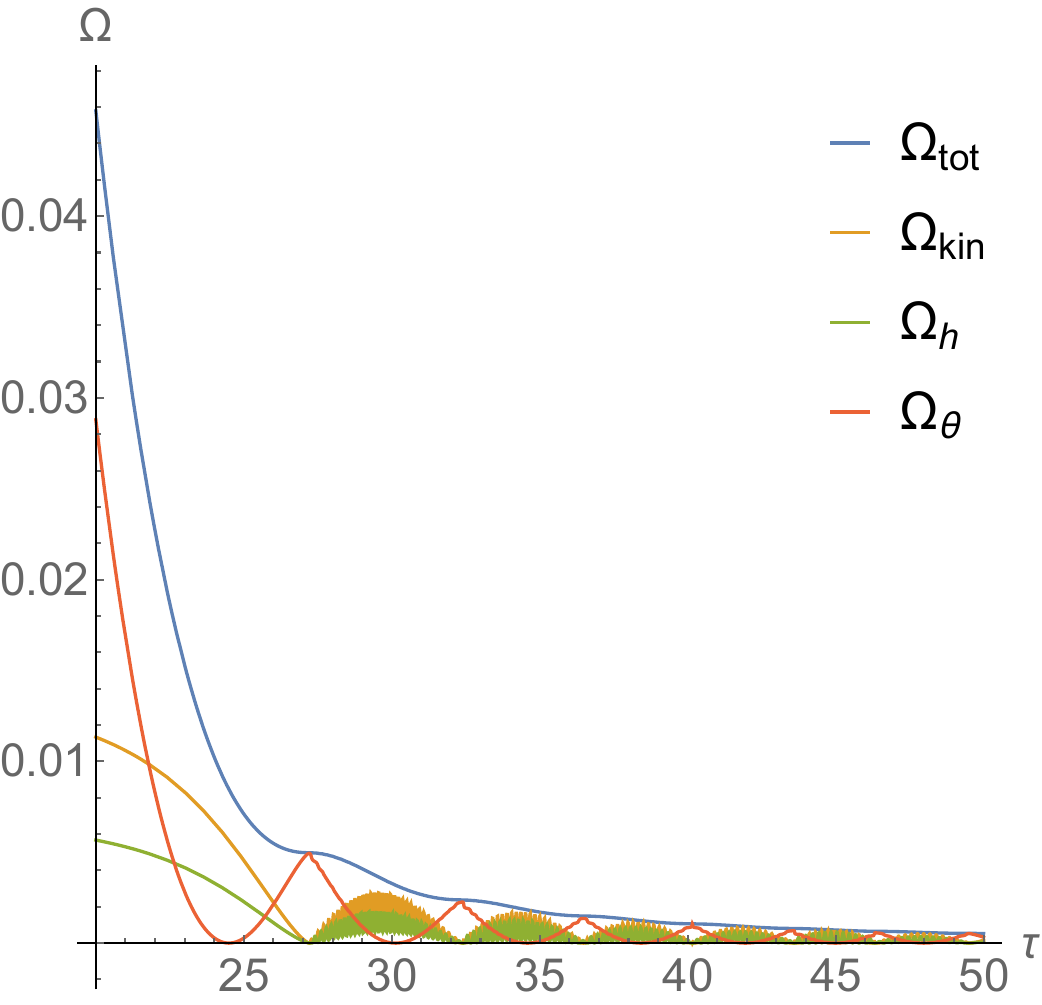}
	\end{center}
	\caption{\label{fig_density_sc}Left-panel: the evolution of the density fraction $\Omega_i = \rho_i /\Lambda_I^4$ with respect to $\tau = H_I \,t$ in the large-mixing case with $h_0 / \Lambda = 67$. The right-panel shows the evolution from $\tau = 20$ for the case with $h_0 / \Lambda = 67$.
	}
\end{figure}
\bigskip\noindent
\textbf{Large mixing.}
In the case of $h_0 \gg \Lambda$, the Higgs mass enhanced by inflaton velocity is so large  that the Higgs oscillations occur near a new minimum $h = h_{\rm new} $ for a short period of time. Given that the energy taken away by the Higgs sector is also important, the kinetic energy $\rho_{\rm kin}$ cannot become dominant at the first stage of preheating. Instead, the inflaton velocity reaches a nearly constant value $\dot{\theta} \approx \sqrt{\lambda}h_{\rm new}$ in a short time as Higgs stablized at the new minimum. This effectively results in an additional phase of slow-roll with $\theta = \theta_0 - \sqrt{\lambda} h_{\rm new} t$ decreases monotonically in time.
Figure \ref{fig_strong-coupling} shows an example with $m_h \gg m_\phi$, where one can see that $H \simeq m_\phi^2\theta/(3h_{\rm new}^2\dot{\theta})$ decreases in proportion to $\theta$ during the additional epoch of slow-roll. 

As the inflaton value decreases, $\theta$ starts to oscillate when $V_{\rm osc}$ becomes comparable to the kinetic energy $\rho_{\rm kin} \simeq h^2\dot{\theta}^2/2$. As shown in Fig.~\ref{fig_density_wc}, the kinetic energy and the Higgs density start to play important roles in the mixed $\theta$-$h$ oscillation for $\tau > 20$, and the background evolution in general deviates from that of the matter domination. 

%\begin{figure}
%	\begin{center}
%		\includegraphics[width=70mm]{density_tau_sc1.pdf}
%		\hfill
%		\includegraphics[width=70mm]{density_tau_sc2.pdf}
%	\end{center}
%	\caption{\label{fig_density_sc}The evolution of the density fraction $\Omega_i = \rho_i /\Lambda_I^4$ with respect to $\tau = H_I \,t$ in the case with $h_0 / \Lambda = 67$. The right-panel shows the evolution from $\tau = 20$.
%	}
%\end{figure}

\bigskip\noindent
\textbf{Shift-symmetry breaking.}
A principal assumption for driving the forced relaxation is that the post-inflationary dynamics is led by Higgs-inflaton couplings which are derivative in $\phi$. Non-derivative couplings, including renormalizable Higgs portal couplings $g_s \Phi_{H}^\dagger\Phi_{H} \phi^2$ and $M  \Phi_{H}^\dagger\Phi_{H} \phi$, may be suppressed during inflation due to the approximated shift symmetry in the inflaton sector. However, the shift symmetry is explicitly broken during preheating and non-derivative couplings in general are non-vanished. The role of non-derivative couplings are clearly model-dependent, and we show that forced relaxation can be realized without the protection of shift symmetry. 

We consider as a typical example the presence of a Higgs portal coupling $g_s \Phi_{H}^\dagger\Phi_{H} \phi^2$, which changes the effective potential as $V_{\rm osc}(\phi) = \frac{1}{2}m_\phi^2 \phi^2 + \frac{1}{2}g_s h^2 \phi^2$. Taking initial conditions in the decoupling limit ($h_0 \ll \Lambda$), the oscillating amplitude can be approximated by $\phi_{\rm osc}^2 \approx \dot{\phi}_{\rm osc}^2/m_\phi^2$. Requiring the dominance of the operator \eqref{toy model} indicates $g_s h^2 \phi_{\rm osc}^2 \ll \dot{\phi}_{\rm osc}^2 h^2/\Lambda^2$, which translates into a constraint $g_s \ll m_\phi^2/\Lambda^2$. We provide in the right panel of Figure \ref{fig_density_wc} a numerical test of the modified system
\begin{align} \label{eom:shift_symm_breaking}
(\Lambda^2+ h^2)\left(\ddot{\phi} +3 H\dot{\phi} \right)+ 2h\dot{h} \dot{\phi}+   m_\phi^2\phi + g_s h^2 \phi&= 0, \\
\ddot{h} + 3H \dot{h} + \lambda h^3 +g_s\phi^2 h&=  h \frac{\dot{\phi}^2}{\Lambda^2}, \label{eom:shift_symm_breaking_h}
\end{align} 
with the extreme value $g_s = m_\phi^2/\Lambda^2$. The density of the portal coupling is not included, so that the definitions of $\Omega_i$ are same as the left panel. The result shows that even the density of the portal coupling $g_s \Phi_{H}^\dagger\Phi_{H} \phi^2$ is not negligible at the beginning of preheating, the derivative operator \eqref{toy model} comes to dominate soon at $\tau \ll 1$ (as reflected by the evolution of $\Omega_{\rm kin}$) and the Higgs relaxation covers a range with $h \gg h_0$. We, therefore, conclude that forced relaxation can be realized by  $g_s \leq m_\phi^2/\Lambda^2$. A similar discussion can be applied to the coupling $M  \Phi_{H}^\dagger\Phi_{H} \phi$, and one obtains a constraint as $M/m_\phi \leq \dot{\phi}_{\rm osc}/\Lambda^2 \approx \Lambda_{I}^2/\Lambda^2$. 
For simplicity, we will not include corrections from non-derivative terms in the following discussion.

\subsubsection{Reheating}

Let us now consider the decay of the inflaton as $H$ approaches to $\Gamma_{I}$. From now on we restrict the discussion to the decoupling limit ($h_0 \ll \Lambda$) which is sufficient for our purpose, yet simplifying the relevant computations significantly.  
Since the Higgs backreaction is negligible in this limit, the $\theta$ oscillation is well approximated by the standard coherent oscillation in the single-field scenario, where we can replace the $\dot{\theta}^{2}$ term in the Eq.~\eqref{eom:h} by its expectation value averaged over an oscillation cycle as
\begin{equation}
\dot{\theta}_{\mathrm{osc}}^{2}=\left.\left\langle \dot{\theta}^{2}\right\rangle \right|_{\mathrm{cycle}}=\rho_{\theta}/\Lambda^{2},\label{eq:theta dot approximation}
\end{equation}
The slow-roll condition $\epsilon \simeq 3\dot{\phi}^2_0/(2\rho_\theta) \ll 1$ implies  $\dot{\theta}_0^2 \ll \dot{\theta}_{\mathrm{osc}}^{2}$.
We explore in Appendix \ref{Appendix} the conditions for the validity of the approximation \eqref{eq:theta dot approximation}.

After the inflaton oscillations are averaged out, the Higgs evolution becomes decoupled from the background. 
The background equations for $\rho_{\theta}$, $H=\dot{a}/a$, and the radiation energy density $\rho_{R}$ are given by
\begin{align}
\dot{\rho}_{\theta}+3H\rho_{\theta}+\Gamma_{I}\rho_{\theta} & =0\label{eq:Inflaton decay}\\
\dot{\rho}_{R}+4H\rho_{R} & =\Gamma_{I}\rho_{\theta}  \label{eq:radiation increase}\\
H^{2} & =\left(\rho_{\theta}+\rho_{R}\right)/3M_{pl}^{2}.   \label{eq:FRW_reheating}
\end{align}
In the matter-dominated background Eq.~\eqref{eq:Inflaton decay} has a solution
\begin{equation}\label{inflaton_density_smoothed}
\rho_{\theta}\left(t\right)=\Lambda_{I}^{4}a^{-3}\left(t\right)e^{-\Gamma_{I}t},
\end{equation}
where $\Lambda_{I}$ is the energy scale of inflation used as our initial condition for $\rho_\theta(t= t_0)=\Lambda^{2}\dot{\theta}_{\mathrm{osc}}^{2}$. Here we assume the coherent oscillations start at $t_0=0$, and the scale factor is normalized as $a\left(0\right)=1$.

With the decay of the inflaton into radiation, the finite temperature effect can become important before or after the time when reheating reaches the maximum temperature, depending on the size of $\Gamma_I$~\cite{Kusenko:2017kdr}, as
\begin{align}
T_{\rm max} \simeq \left(\frac{1}{\sqrt{8\pi}g_\ast} \Lambda_I^2\Gamma_IM_p \right)^{1/4},
\end{align}
%$T_{\rm max} \sim (\Lambda_I^2\Gamma_IM_p/(\sqrt{8\pi}g_\ast))^{1/4}$~\cite{Kusenko:2017kdr} 
where the effective massless degrees of freedom $g_\ast = 106.75$ for a temperature larger than $300$ GeV. 
The finite-temperature correction to the Higgs potential is contributed as a positive thermal mass term 
\begin{equation}\label{thermal mass}
V_{T}\simeq\frac{1}{2}\alpha_{T}^{2}T^{2}h^{2},
\end{equation}
where $\alpha_{T}\simeq0.33$ at energy scale about $10^{13}\,\text{GeV}$, and $T$ is the temperature of the plasma. Therefore, the Higgs has an effective potential 
\begin{equation}
V=\frac{1}{4}\lambda h^{4}-\frac{1}{2}\left(\frac{\rho_{\theta}}{\Lambda^{2}}-\alpha_{T}^{2}T^{2}\right)h^{2}
\end{equation}
with a time-dependent equilibrium minimum at 
\begin{equation}
h_{\min}=\begin{cases}\left[\frac{ \rho_{\theta}/\Lambda^{2} -\alpha_{T}^{2}T^{2}}{\lambda}\right]^{1/2}
& \text{if }\qquad \rho_{\theta} > \alpha_{T}^{2}T^{2}\Lambda^{2}, \\
0 & \text{if }\qquad \rho_{\theta} \leq \alpha_{T}^{2}T^{2}\Lambda^{2}.
\end{cases}
\end{equation}
Note that the reheating temperature can be obtained as a function of time via $T(t)\approx [30\rho_R(t)/(\pi^2 g_\ast)]^{1/4}$, where $T \sim t^{-1/2}$ as reheating completed.

\begin{figure}
	\begin{center}
		\includegraphics[width=12cm]{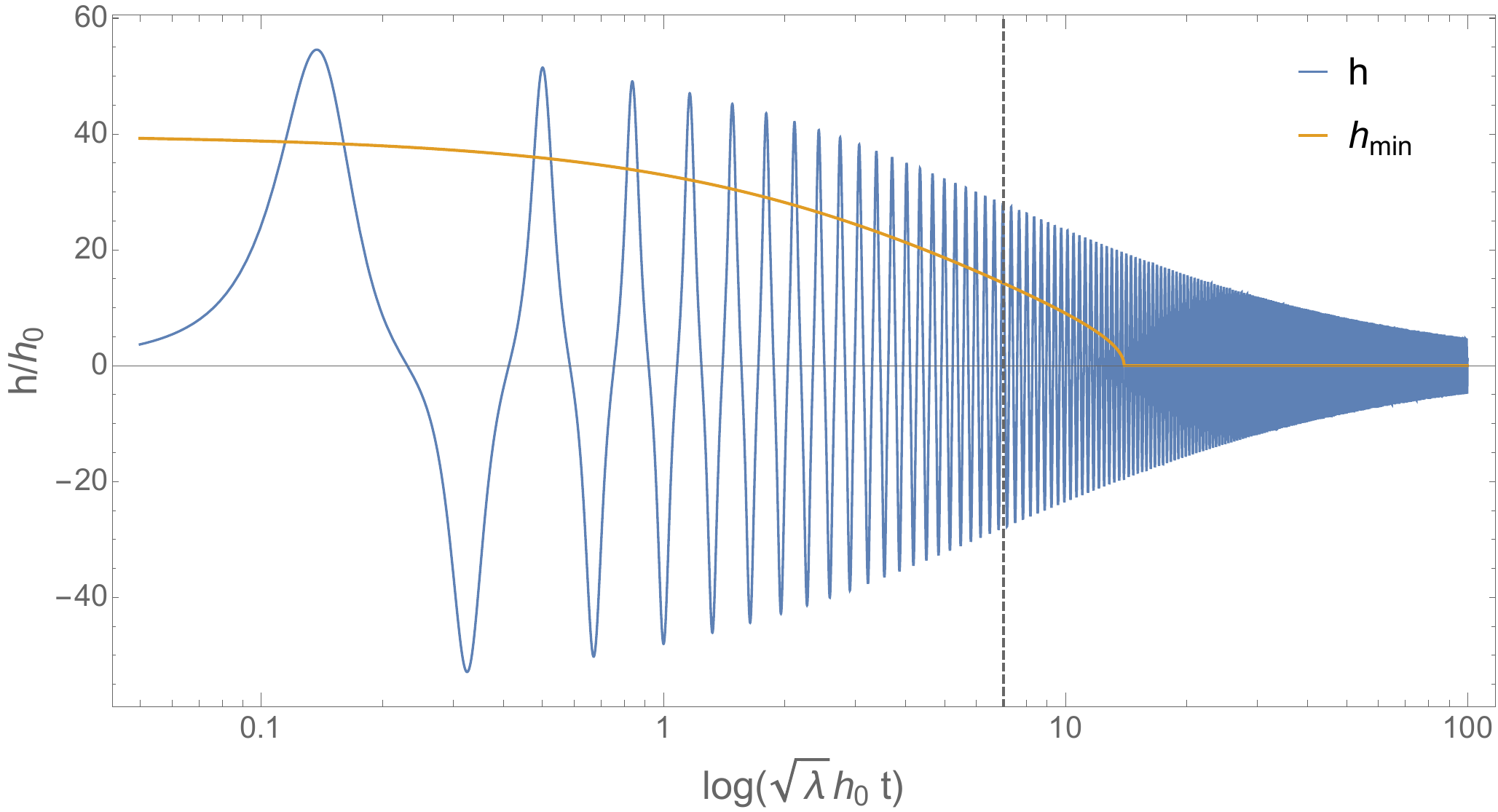} 
		%\par
	\end{center}
	\caption{Higgs relaxation from forced initial conditions. The parameters are $\Lambda=10^{16}\,\text{GeV}$, $\Lambda_{I}=2\times10^{14}\,\text{GeV}$, $\Gamma_{I}=10^{6}\,\text{GeV}$, and $\lambda=0.001$. The blue line shows the evolution of the Higgs VEV during reheating. The yellow line shows the positive minimum of the Higgs potential. The dashed line indicates the time when maximal temperature is reached. \label{fig:Higgs-evolution}}
\end{figure}

By solving the evolution of the background $\rho_{\theta}\left(t\right)$ and $a\left(t\right)$, we can then find out the dynamics of the Higgs VEV by the modified equation of motion
\begin{equation}\label{eq:eom_h_finiteT}
\ddot{h}+3H\dot{h}+\lambda h^{3}-\left(\frac{\rho_{\theta}}{\Lambda^{2}}-\alpha_{T}^{2}T^{2}\right)h=0,
\end{equation}
with the initial Higgs VEV $h_{0}=\dot{\theta}_{0}/\sqrt{\lambda}$.
Figure~\ref{fig:Higgs-evolution} shows a typical example of the Higgs relaxation from forced initial conditions where the oscillating amplitudes can reach $h \gg h_0$. We see that the finite temperature effect dominates the Higgs potential after the maximum reheating temperature is reached. Note that the decay of Higgs into SM particles via the non-perturbative channel of parametric resonance and the perturbative channel of thermalization are not taken into account 
{\footnote{The non-perturbative and perturbative decay of Higgs into SM particles are not efficient before the maximal reheating temperature is reached, see \cite{Yang:2015ida} for the case of free-fall relaxation and Appendix~\ref{AppendixC} for the case of forced relaxation.}}
{\footnote{Parametric resonance of the Higgs mode functions does not necessarily happen in forced relaxation by a proper choice of the inflaton mass $m_\phi$, see Appendix~\ref{AppendixB}.}.}

\section{Leptogenesis from forced relaxation }\label{Sec. leptogenesis}

In this section we discuss leptogenesis from forced relaxation, building on the idea of scalar field relaxation leptogenesis~\cite{Kusenko:2014lra,Pearce:2015nga,Yang:2015ida,Kusenko:2017kdr}. In the framework of relaxation leptogenesis, the lepton/baryon asymmetry is sourced by the classical motion of a scalar field in the post-inflationary epoch. In the conventional setup \cite{Kusenko:2014lra,Pearce:2015nga,Yang:2015ida}, one uses Higgs as the source field to generate the lepton number asymmetry where the initial Higgs VEV $h_0$ is the maximal value of the relaxation process.
%attributed to the accumulation of long-wavelength quantum fluctuations during inflation as $h_0 \equiv \sqrt{\langle h^2 \rangle}$. Since the Higgs potential is very shallow at very high energy, the Higgs relaxation begins with a value $h_0\sim H_I/(\lambda)^{1/4}$ from the end of inflation.
In our case, the initial Higgs VEV is spontaneously generated by the  symmetry breaking due to the Higgs-inflaton coupling $\mathcal{O}_{\phi h}$. For the model \eqref{toy model} $h_0$ is controlled by the parameter $\dot{\theta}_{0}$. The Higgs relaxation at the end of inflation is dynamically triggered by the sudden increase of the inflaton kinetic energy. 

\bigskip\noindent
\textbf{External chemical potential.}
To realize leptogenesis from the post-inflationary Higgs motion, one introduces the derivative coupling between the Higgs field and the $B+L$ fermion current:
\begin{equation}
\mathcal{O}_{6}=\frac{-2}{\Lambda_{n}^{2}}\Phi_{H}^{\dagger}\Phi_{H}\partial_{\mu}j_{B+L}^{\mu}, \label{eq:O6-1}
\end{equation}
where $\Lambda_{n}$ is the energy scale above which this operator becomes irrelevant. The form of this operator is familiar in the electroweak anomaly equation as $\partial_\mu j^\mu \sim -g^2 W\tilde{W} + g^{\prime 2} A\tilde{A}$, where $W$ and $A$ are $SU_L(2)$ and $U_Y(1)$ gauge fields, respectively. The cutoff scale can come from integrating out heavy states with mass scale $\Lambda_{n} = M_{n}$ or from the thermal loop corrections with a temperature $\Lambda_{n} = T$. One can move the derivative from the fermion current $j_{B+L}^{\mu}$ to the Higgs field $h$ via integration by part and finds
\begin{equation}
\mathcal{O}_{6}=\frac{1}{\Lambda_{n}^{2}}\left(\partial_{\mu}h^{2}\right)j_{B+L}^{\mu}. \label{eq:O6}
\end{equation}
For a patch of the universe where the Higgs field $h$ is approximately homogeneous, this operator describes an external source of the $B+L$ charge density due to the time derivative of the Higgs field,
\begin{equation}
\mathcal{O}_{6}=\frac{1}{\Lambda_{n}^{2}}\left(\partial_{0}h^{2}\right)\rho_{B+L}. \label{eq:O6-0}
\end{equation}
This is an operator that violates the charge, parity and time reversal (CPT) invariance.
To see this, it is useful to treat such an external source as an effective chemical potential during the thermal equilibrium of the $B+L$ fermions as
\begin{equation}
\mu_{\rm eff}=-\frac{\partial_{0}h^{2}}{\Lambda_{n}^{2}}. \label{eq:chemical potential}
\end{equation}
The chemical potential $\mu_{\rm eff}$ shifts the ground state energy of antiparticles (or particles) by $ \partial_{0}h^2/\Lambda_{n}^2$ (or $ -\partial_{0}h^2/\Lambda_{n}^2$) at thermal equilibrium \cite{Yang:2015ida}.
Before the average Higgs VEV decreases to zero, the chemical potential $\mu_{\rm eff}$ will favor the production of anti-lepton over lepton, resulting in a net lepton number $n_L \sim \mu_{\rm eff} T^2$ through lepton-number-violating processes \cite{Cohen:1987vi,Kusenko:2014lra}. The sphaleron process then redistributes the lepton number into baryon number, leading to a successful baryogenesis.

\begin{figure}
	\begin{centering}
		\includegraphics[width=0.85\columnwidth]{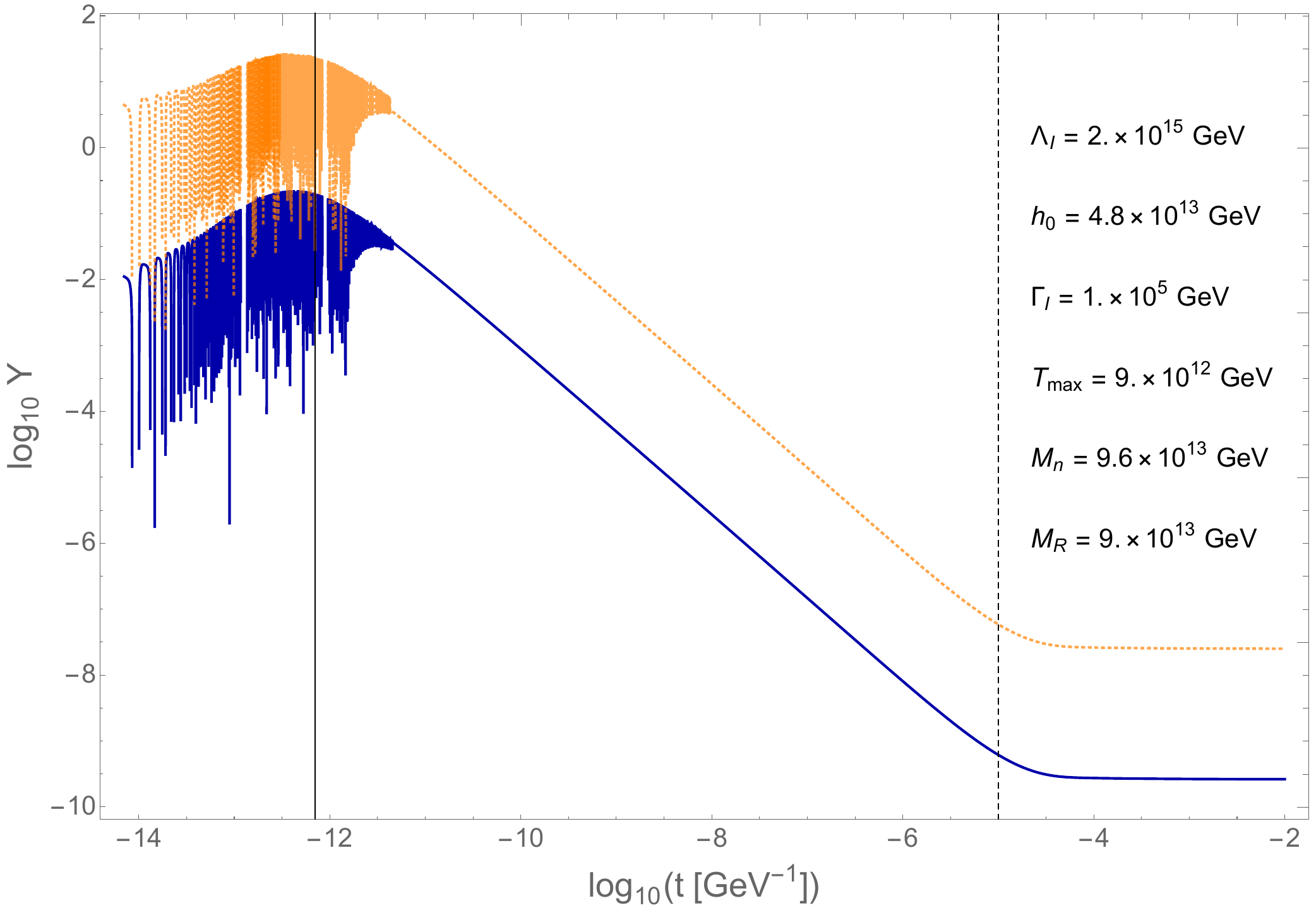}
		\par\end{centering}
	\caption{The lepton asymmetry from forced relaxation. The blue-solid line shows an example with $\Lambda_{n} = M_n = 2 h_0$, and the orange-dotted line shows an example with $\Lambda_{n} = T$. The vertical solid line indicates the time when the maximum temperature is reached, and the vertical dashed line indicates the beginning of radiation dominated epoch. 
		\label{fig:lepton_asymmetry}}
\end{figure}

\bigskip\noindent
\textbf{Lepton asymmetry.}
We consider the typical lepton-number-violating process from a family of right-handed neutrinos $N_{R,i}$ with Majorana mass terms $M_{R,i}$ that can explain the smallness of the observed neutrino masses by the (Type-I) see-saw mechanism. From observations we fix the neutrino mass $M_\nu \approx 0.1$ eV, and the Yukawa coupling is given by
\begin{align}
y_\nu = \left(\frac{2M_\nu M_R}{v_{\rm EW}^2}\right)^{1/2},
\end{align}
%\begin{equation}
%\frac{y_{\nu}^{2}v_{EW}^2}{2M_{R}}=M_{\nu}\sim0.1\,\text{eV},
%\end{equation}
where $v_{\rm EW}=246.22\,\text{GeV}$ is the Higgs VEV at the electroweak minimum, and $M_R$ corresponds to the lightest right-handed neutrino mass in the family. The Yukawa coupling $y_\nu$ is chosen with perturbativity $y_\nu^2/(4\pi) < 1$.
We shall require $M_R > m_h$ and $M_R > T_{\rm max}$ to prevent production of right-handed neutrinos from the decay of Higgs condensate or the thermal plasma. These requirements suppress leptogenesis from the thermalization channel \cite{Fukugita:1986hr}. Examples for lepton-number-violating processes involved with the exchange of the heavy right-handed neutrino can be found in \cite{Kusenko:2014lra,Yang:2015ida}.

The evolution of the lepton number density $n_L = n_\nu - n_{\bar{\nu}}$ towards equilibrium in the detailed balance regime is illustrated by a system of Boltzmann equations including the external chemical potential \eqref{eq:chemical potential}. To leading order in $\mu_{\rm eff}/T$ as $\langle \partial_{0}h^2\rangle$ approaches zero, the Boltzmann equation can be approximated by
\begin{align}\label{boltzmann leading}
\dot{n}_L + 3H n_L = -2 \left\langle\sigma v\right\rangle n_0^{\rm eq} \left(n_L - 2 \frac{\mu_{\rm eff}}{T}n_0^{\rm eq} \right),
\end{align}
where $n_0^{\rm eq} \equiv T^3/\pi^2$ is the equilibrium number density of left-handed neutrinos in the limit of $\mu_{\rm eff} =0$.
The thermally averaged cross section $\left\langle\sigma v\right\rangle$ includes at least the processes from $\bar{\nu}_L h \leftrightarrow \nu_L h$ and $\nu_L \nu_L \leftrightarrow h\,h$ with the assumption that the reaction rate is the same for the reversed processes. These reaction rates are computed in the limit of $\mu_{\rm eff} =0$, where the deviation from the actual values with a finite $\mu_{\rm eff}$ is higher-order in $\mu_{\rm eff}/T$.

We show in Figure \ref{fig:lepton_asymmetry} some numerical examples in the decoupling limit with $\Lambda = 2\times 10^{15}$ GeV and different choices of $\Lambda_{n}$. The final lepton asymmetry is given by the ratio
\begin{align}
Y_L(t) = \frac{n_L(t)}{s(t)} = \frac{45}{2\pi^2 g_\ast}\frac{n_L(t)}{T^3(t)},
\end{align}
where $s = 2\pi^2 g_\ast T^3/45$ is the entropy density.
At the beginning of reheating, $Y_L(t)$ is produced and washed out due to the rapid oscillation of the chemical potential $\mu_{\rm eff}$ until $T$ reaches the maximal temperature. As the temperature decreases from the maximal value, the cross section becomes too small so that $n_L$ can freeze to a non-zero value before $\mu_{\rm eff}$ vanishes. Once reheating completes where the Universe becomes radiation domination ($T\sim a^{-1}$), $Y_L(t)$ frozen out to a constant value. Note that $\mu_{\rm eff}$ gradually vanishes as the finite temperature effect dominates the Higgs potential where $h_{\rm min} = 0$.

\begin{figure}
	\begin{centering}
		\includegraphics[width=0.8\columnwidth]{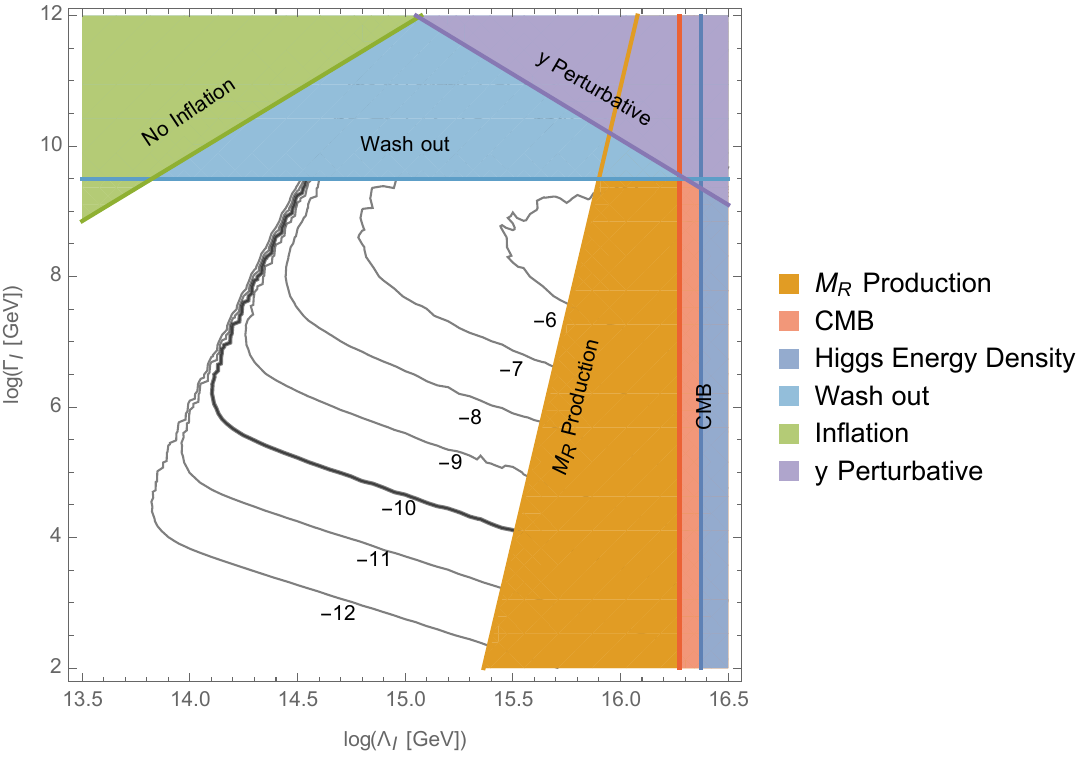}
		\par\end{centering}
	\caption{The parameter space for forced relaxation with respect to inflaton decay width, $\Gamma_{I}$, and the scale of inflation, $\Lambda_{I}$. The contours denote the logarithmic power of the final lepton asymmetry $\left(n = \log_{10} Y_{L} \right)$. The shaded region is the parameter space excluded by various kinds of constraints.
		\label{fig:The-parameter-space}}
\end{figure}

\bigskip\noindent
\textbf{Parameter space.}
For the choice of the cutoff $\Lambda_n = M_n$ as an independent model parameter, $\mu_{\rm eff}\propto M_n^{-2}$ so that the final lepton density $n_L \sim \mu_{\rm eff} T^2$ is scaling with the change of $M_n$. This allows us to find a wide range of parameter space for the desired asymmetry $Y_L \sim 10^{-10}$ by a rescaling of $M_n$. Note that the example in Figure~\ref{fig:lepton_asymmetry} satisfies $M_n > h_0$ and $M_n > T_{\rm max}$, which justifies $\mathcal{O}_6$ as a well-controlled higher-dimensional operator in the effective field theory. It is therefore an important advantage to consider leptogenesis from forced-relaxation. On the other hand, the conditions $M_n > h_0$ and $M_n > T_{\rm max}$ are found to put a strict constraint in the parameter space to find $Y_L \sim 10^{-10}$ from the free-fall initial conditions \cite{Yang:2015ida}. 

The parameter space for the desired lepton asymmetry can be narrowed down by taking $\Lambda_{n} = T$ as a cutoff fixed by the reheating process. Figure~\ref{fig:The-parameter-space} shows the scan of parameter space for the final lepton asymmetry $Y_L$ with respect to the inflaton decay width $\Gamma_{I}$ and the scale of inflation $\Lambda_{I}$. In this figure model parameters are set to $\Lambda=10^{16}\,\text{GeV}$, $\lambda=0.001$, and $\Lambda_{n}=T$. With $\dot{\theta}_0$ given by the power spectrum \eqref{power_spectrum}, the condition $\Lambda_{I}^4 > \lambda h_0^4/4$ for a subdominant Higgs density leads to
\begin{align}
\Lambda > \frac{1}{6\pi\sqrt{2A_s \lambda^{1/2}}}\frac{\Lambda_{I}^3}{M_p^2}.
\end{align} 
This condition is incorporated by the CMB observations on the slow-roll parameter $\epsilon_1\equiv -\dot{H}/H^2 \approx \dot{\phi}^2/\Lambda_{I}^4$ in the single-field inflation with $\epsilon_1 <0.0068$ \cite{Ade:2015lrj,Akrami:2018odb}. To avoid entering into the regime of thermal leptogenesis \cite{Fukugita:1986hr} via the production of right-handed neutrinos from the decay of the Higgs condensate, we impose the condition $M_R > h_0$. We use a conservative choice $M_R = 10 T_{\rm max} $ to avoid $N_R$ production, such that the condition $M_R > h_0$ implies
\begin{align}
\Lambda_{I} < \left(60\pi \sqrt{\lambda A_s}\right)^{2/7} \left(\frac{1}{\sqrt{8\pi} g_\ast}\right)^{1/14} M_p^{9/14}\Lambda^{2/7} \Gamma_{I}^{1/14},
\end{align}
 For $\Gamma_I \gtrsim 2\times10^{9}\,\text{GeV}$ (the ``wash out'' region in Figure~\ref{fig:The-parameter-space}), the resulting asymmetry keeps being washed out by the Higgs oscillation even after reheating. The ``$y$ perturbative'' region in upper right corner indicates the parameter space where the Yukawa coupling of neutrinos is not perturbative. For $3H_I \lesssim \Gamma_I$, no inflation happens, which is denoted as ``no inflation'' region in the upper left corner.

%The yellow shaded region denotes the constraint on $\Lambda_I$ according to the Planck observations. The constraint on the inflaton kinetic energy~($\epsilon_1 <0.0068$ \cite{Ade:2015lrj,Akrami:2018odb}) gives a lower bound on the $\Lambda_I$, shown as the red region. 
%\section{Signatures in primordial non-Gaussianities}

\section{Conclusions}\label{Sec. conclusion}
In this work we investigated leptogenesis due to relaxation of the Higgs condensate from the initial vacuum expectation value (VEV) developed during inflation to the electroweak minimum. We focus on inflationary scenarios that generate a sizable VEV, around or greater than the scale of the Hubble horizon, via spontaneous symmetry breaking in the Higgs potential. As concrete examples for symmetry breaking, we have considered the Higgs evolution during reheating led by the wrong-sign non-minimal coupling with gravity \eqref{eom_free:h} and the higher-dimensional kinetic coupling with inflaton \eqref{eom:h}.

We have identified a new class of post-inflationary relaxation dynamics from the kinetic coupling model \eqref{toy model}, where the inflaton produces an effective external force that pushes the Higgs field away from the origin. The Higgs VEV during forced relaxation receives additional energy from the inflaton sector, so that the oscillation amplitude can be much greater than the initial VEV (that is, $h/h_0 \gg 1$). We have explored the scalar-relaxation leptogenesis in the case of  forced relaxation dynamics and found that the desired final lepton asymmetry can be realized for a wide range of parameters. 

It is remarkable that inflation with a symmetry breaking in the SM sector in general exhibits observable signatures in primordial non-Gaussianities \cite{Kumar:2017ecc,Wu:2018lmx}, which opens a unique window for testing initial conditions of leptogenesis from forced relaxation. These non-Gaussian signatures are generated through the quantum interference between inflaton and the heavy SM fields under the symmetry structure of the inflationary spacetime \cite{Arkani-Hamed:2015bza,Arkani-Hamed:2018kmz,Lee:2016vti,Meerburg:2016zdz,Chen:2016hrz,Chen:2016uwp}, which is exemplified by the so-called quasi-single field inflation \cite{Chen:2009we,Chen:2009zp,Noumi:2012vr,Wang:2018tbf}.  
Unfortunately, the corresponding non-Gaussian signals of leptogenesis from forced relaxation in the decoupling limit ($h_0/\Lambda \ll 1$) demonstrated in this paper are too small to be observationally interesting, since the size of non-Gaussianity is proportional to the Higgs-inflaton couplings. However, the non-Gaussian signals of heavy SM gauge fields can be substantially enhanced by the presence of large gauge-inflaton mixings~\cite{Wu:2018lmx,An:2017hlx,Iyer:2017qzw}. The viability of leptogenesis from the  initial conditions in the large-mixing regime ($h_0/\Lambda \geq 1$) is left for future studies.

\acknowledgments
The authors thank Kohei Kamada, Misao Sasaki, Toyokazu Sekiguchi, Yusuke Yamada, Jun'ichi Yokoyama, and Zhong-Zhi Xianyu for the helpful comments and discussions.
Y.P.W is supported by JSPS International Research Fellows and JSPS KAKENHI Grant-in-Aid for Scientific Research No. 17F17322, and this project was in part supported by Ministry of Science and Technology (MoST) Postdoctoral Research Abroad Program (PRAP) MoST-105-2917-I-564-022.
The work of A.K. was supported by the U.S. Department of Energy (DOE) Grant No. DE-SC0009937 and by the World Premier International Research Center Initiative (WPI), MEXT, Japan.  Part of this work was performed at the Aspen Center for Physics, which is supported by National Science Foundation grant PHY-1607611. 

\appendix

\section{Resonance in forced relaxation}\label{Appendix}
Our numerical evaluation of the lepton asymmetry in Section \ref{Sec. leptogenesis} relies on an important approximation where the inflaton oscillation can be averaged out to become a smoothed background. In reality, both Higgs and inflaton are oscillating so that resonance could happen and develop instability to the system. In this section we investigate the limitation of the smoothing process \eqref{eq:theta dot approximation}, where the (smoothed) initial velocity is only fixed by one parameter as $\dot{\phi}(t_0) = \Lambda_{I}^2$. 

\begin{figure}
	\begin{center}
		\includegraphics[width=76mm]{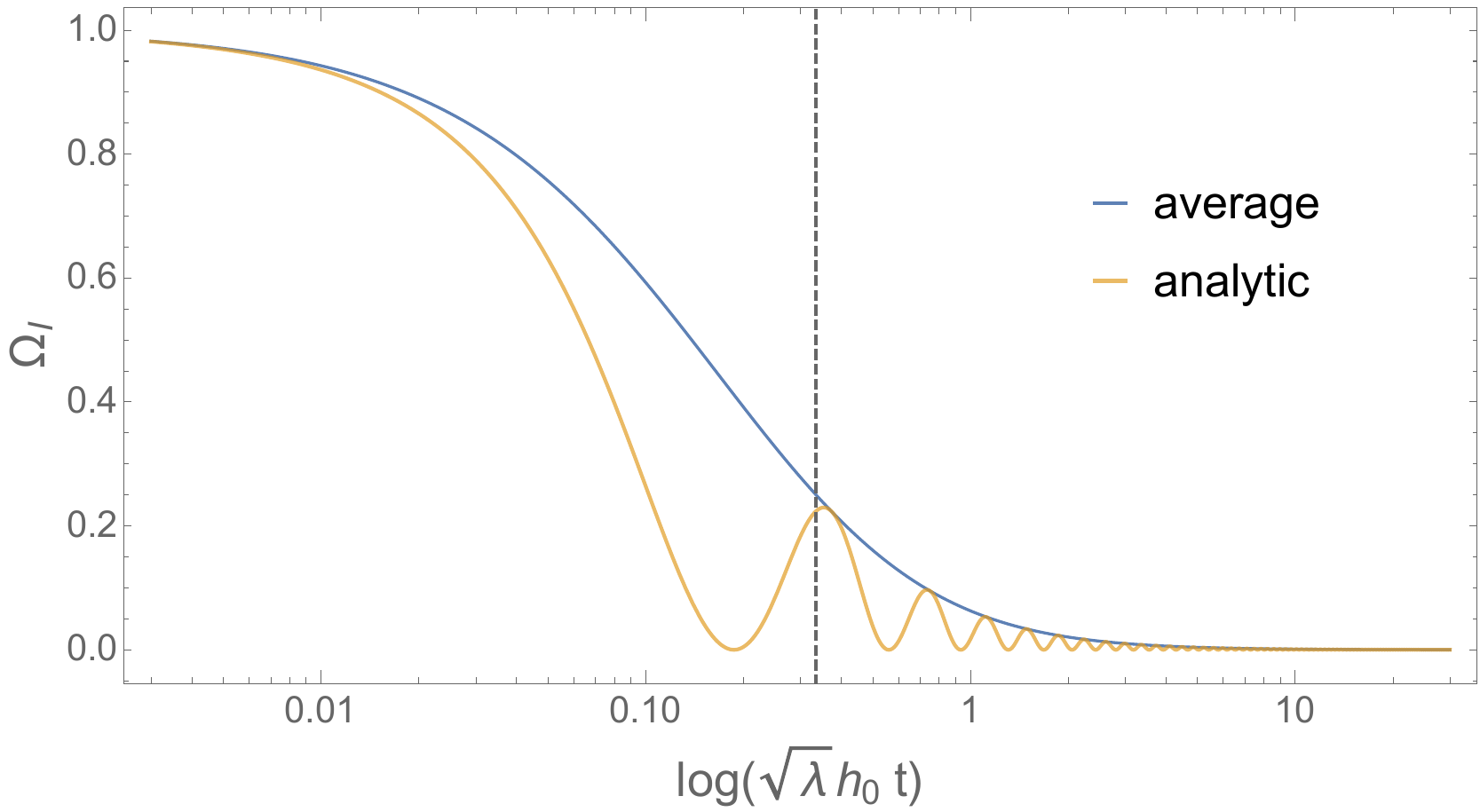}
		\hfill
		\includegraphics[width=76mm]{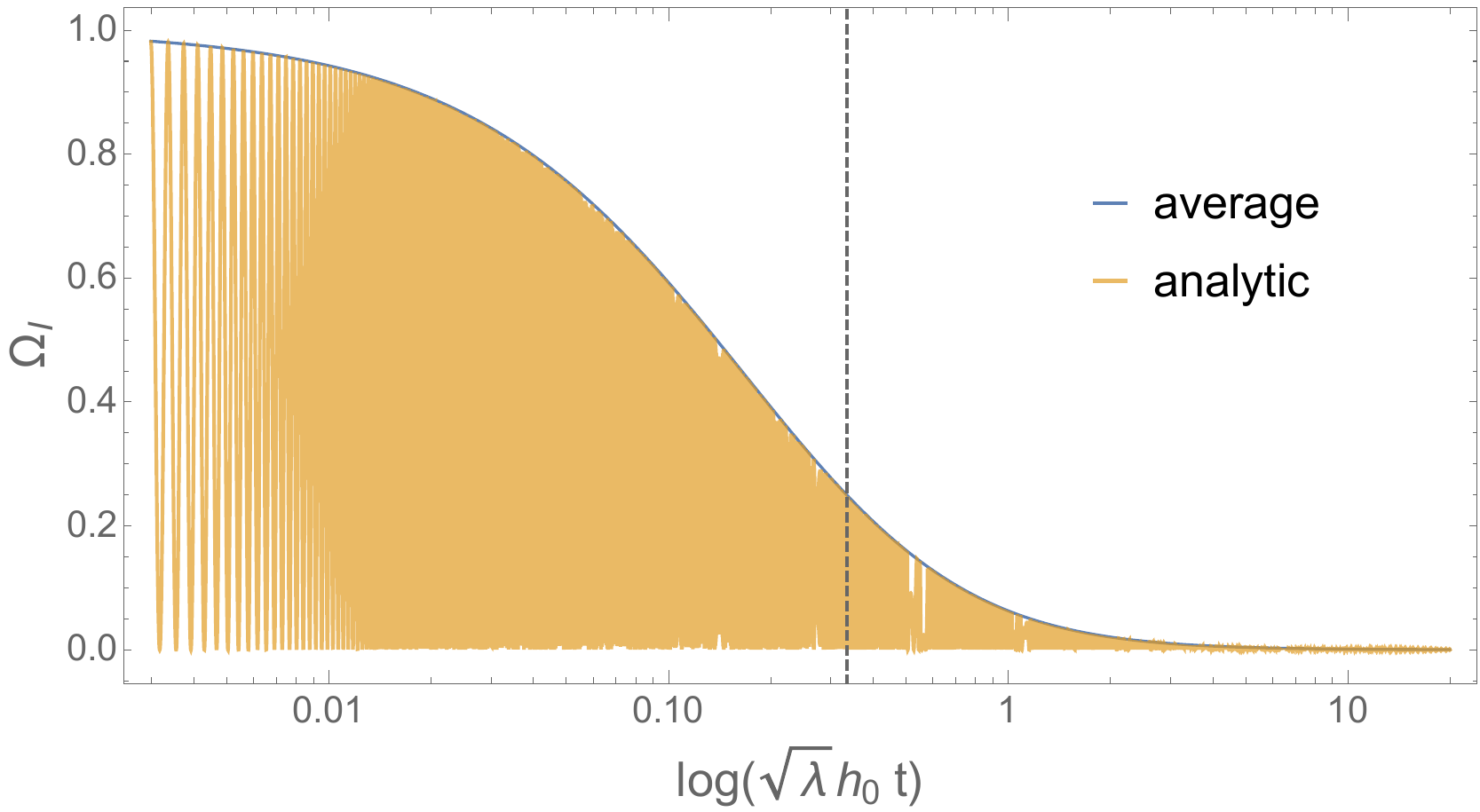}
	\end{center}
	\caption{\label{fig_analytic_density}The evolution of the inflaton density with the mass parameter $m_\phi = 0.01 \times \Lambda_{I}^2/\Lambda$ (left panel) and $m_\phi = 10 \times \Lambda_{I}^2/\Lambda$ (right panel). The blue-solid line shows the averaged expectation value. The dashed line is the time when maximum temperature is reached. 
	}
\end{figure}

\begin{figure}
	\begin{center}
		\includegraphics[width=76mm]{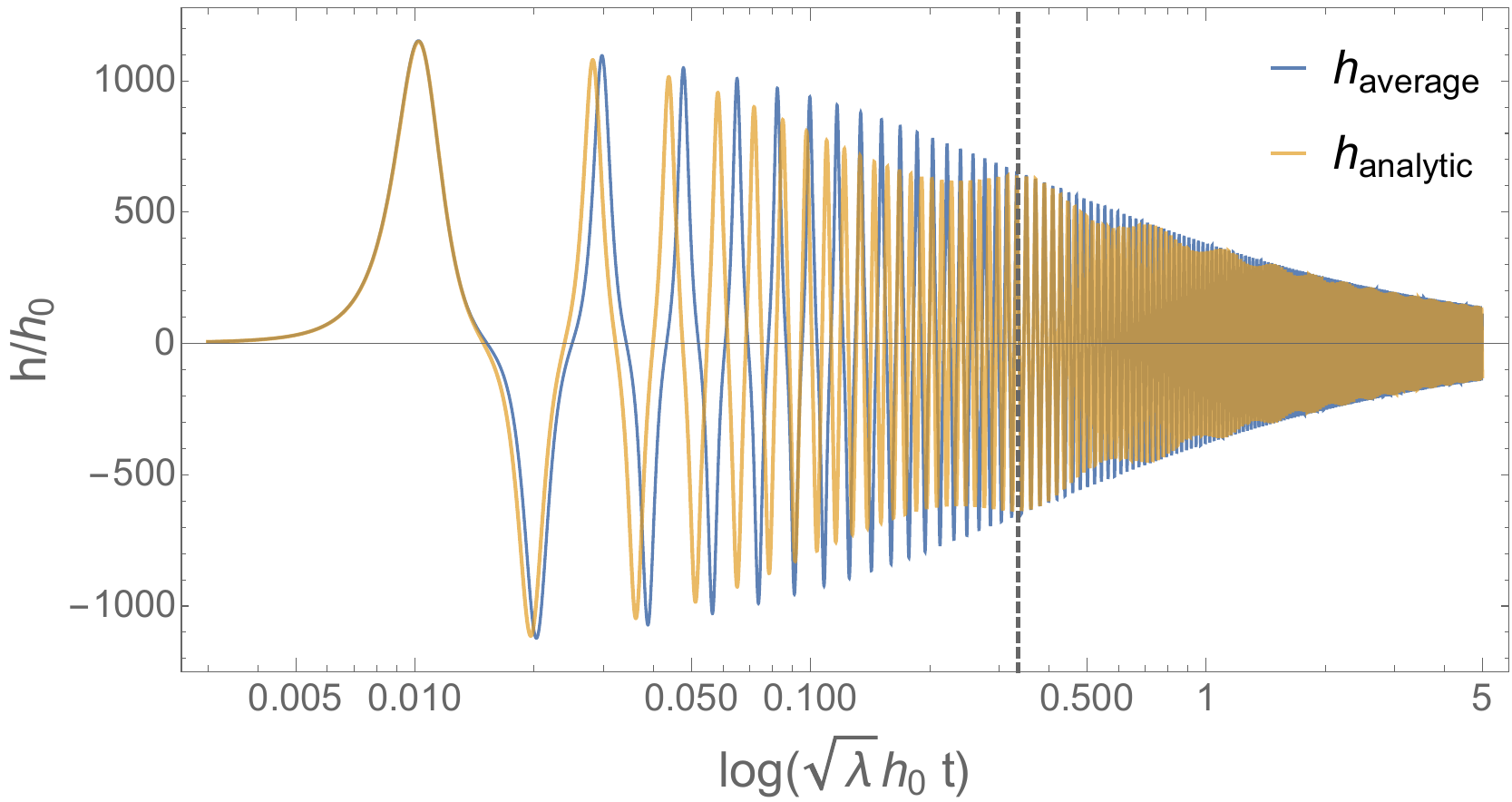}
		\hfill
		\includegraphics[width=76mm]{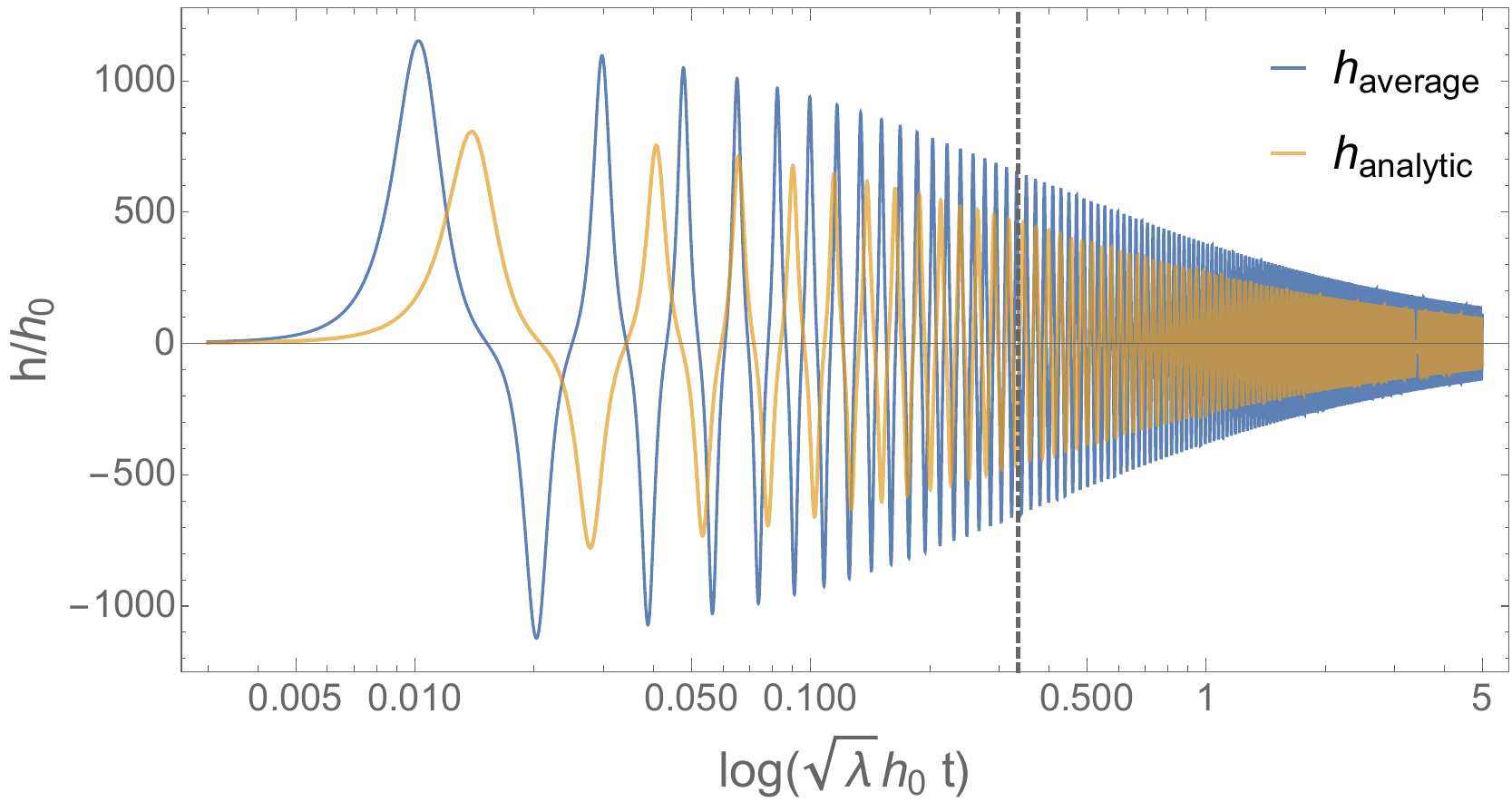}
	\end{center}
	\caption{\label{fig_analytic_h}The evolution of Higgs with the mass parameter $m_\phi = 0.01 \times \Lambda_{I}^2/\Lambda$ (left panel) and $m_\phi = 10 \times \Lambda_{I}^2/\Lambda$ (right panel). The blue-solid line shows the result from a smoothed inflaton density. The dashed line is the time when maximum temperature is reached. 
	}
\end{figure}

The full numerical calculation in the section of preheating has confirmed that the overall expansion rate is $a^{-3}$ (matter-domination like) in the decoupling limit.  We thus describe the harmonic oscillations of the inflaton in the form 
\begin{align}
\phi = \phi_{\rm max} \left(\frac{a}{a_0}\right)^{-3/2} e^{-\Gamma_{I}t/2} \sin\left(m_\phi t + \beta\right),
\end{align}
where $\beta $ is a phase parameter. This analytic form does not capture the exact motion but is sufficient for the study of the resonance. For a given inflaton mass $m_\phi$, the energy density $\rho_{\phi} = \dot{\phi}^2$ is given by
\begin{align}\label{eq:inflaton_analytic}
\rho_\phi^{\rm analytic} \approx m_\phi^2 \phi_{\rm max}^2 \left(\frac{a}{a_0}\right)^{-3} e^{-\Gamma_{I}t}\cos^2(m_\phi t + \beta),
\end{align}
where we only keep the leading term with $m_\phi \gg H_I$ and $\phi_{\rm max}$ can be obtained from the initial condition $\Lambda_{I}^4 = m_\phi^2 \phi_{\rm max}^2 \cos^2(m_\phi t_0 + \beta)$. Figure \ref{fig_analytic_density} shows that the smoothed density $\Omega_I \equiv \rho_\phi/\Lambda_{I}^4 = a^{-3}e^{-\Gamma_{I} t}$ can incorporate the analytic form with a wide range of the inflaton masses. To search the resonance effect, we replace the smoothed density in \eqref{eq:eom_h_finiteT} by the oscillating form as
\begin{equation}\label{eq:eom_h_analytic}
\ddot{h}+3H\dot{h}+\lambda h^{3}-\left(\frac{\rho_{\phi}^{\rm analytic}}{\Lambda^{2}}-\alpha_{T}^{2}T^{2}\right)h=0.
\end{equation}
Having in mind that the Higgs mass scale is around $\Lambda_{I}^2/\Lambda$ at the onset of reheating, we first explore the Higgs dynamics in the cases of $m_\phi \ll \Lambda_{I}^2/\Lambda$ and $m_\phi \gg \Lambda_{I}^2/\Lambda$ in the left and right panels of Figure \ref{fig_analytic_h}, where $\beta = t_0 = 0$ is used. The results indicate that numerical computation with a smoothed inflaton background is a good approximation for $m_\phi \ll \Lambda_{I}^2/\Lambda$ and the real amplitude for $m_\phi \gg \Lambda_{I}^2/\Lambda$ may be smaller due to the rapid oscillation of $\phi$.

\begin{figure}
	\begin{center}
		\includegraphics[width=76mm]{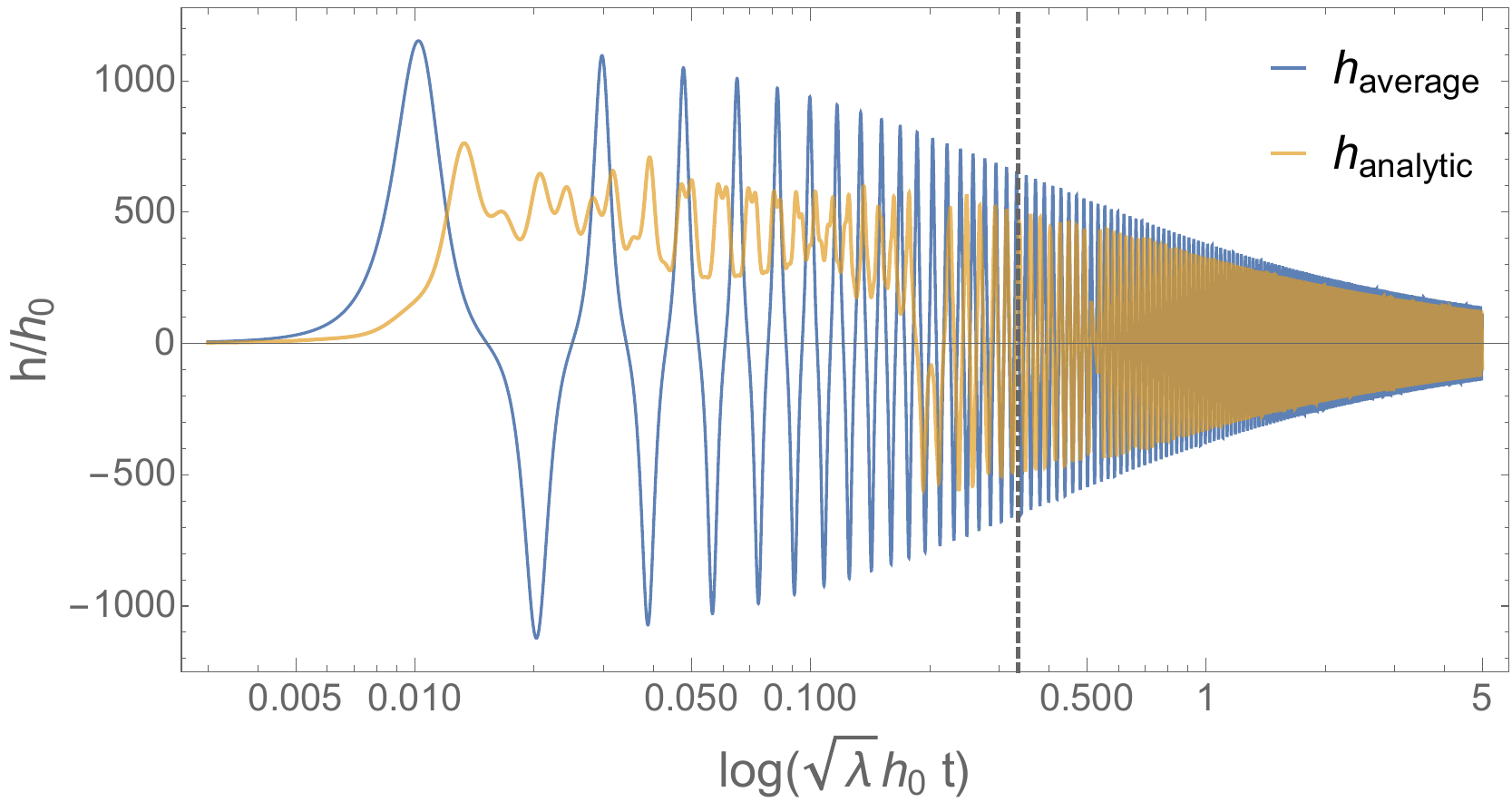}
		\hfill
		\includegraphics[width=76mm]{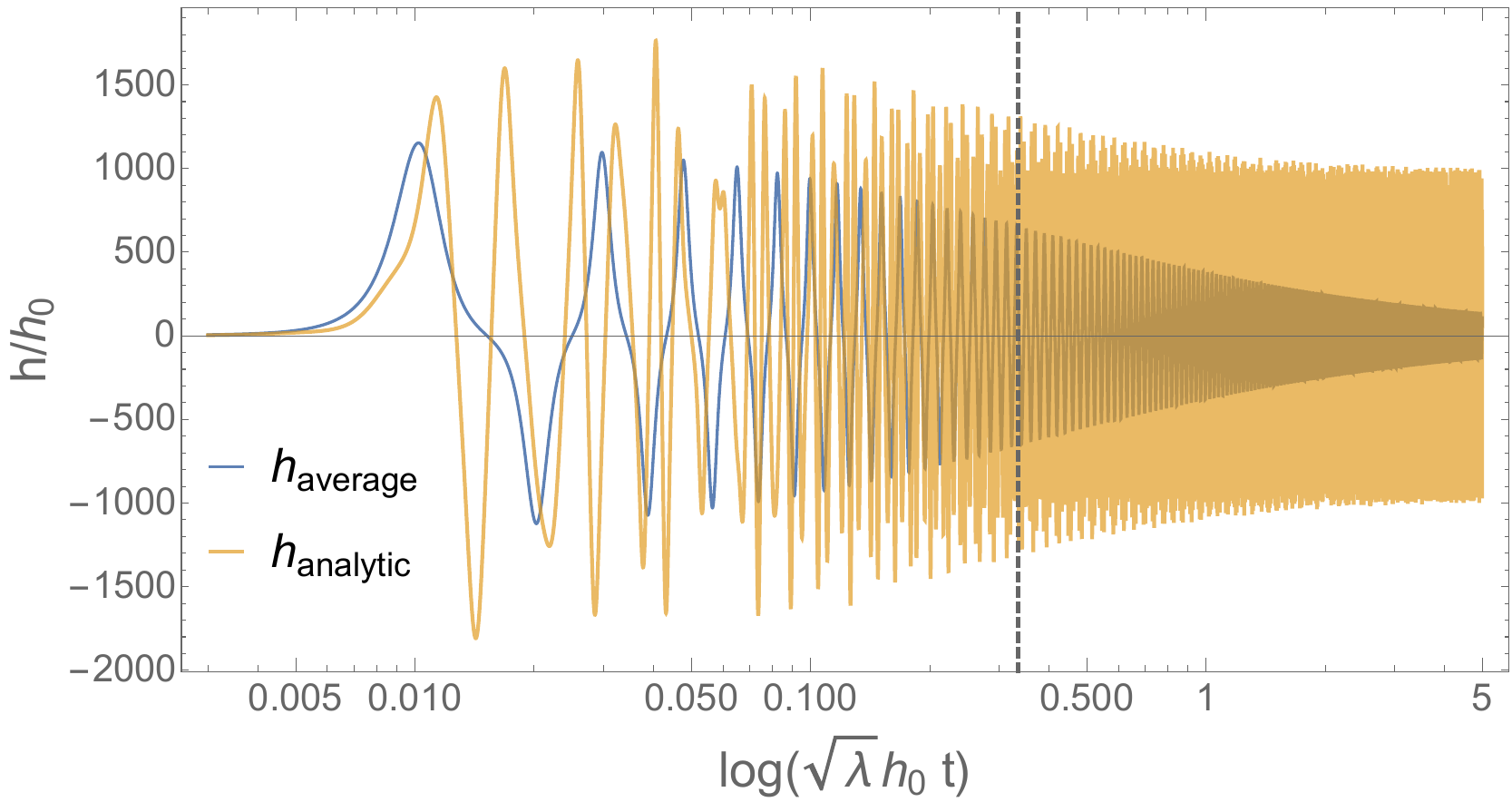}
	\end{center}
	\caption{\label{fig_analytic_h_critical}The evolution of Higgs with the mass parameter $m_\phi = \Lambda_{I}^2/\Lambda$. The phase is chosen as $\beta = 0$ (left panel) and $\beta = \pi/4$ (right panel). The blue-solid line shows the result from a smoothed inflaton density. The dashed line is the time when maximum temperature is reached. 
	}
\end{figure}

We show in Figure \ref{fig_analytic_h_critical} the cases with $m_\phi = \Lambda_{I}^2/\Lambda$ and demonstrate that the resonance can happen instructively or destructively, depending on the phase $\beta$. A scan of the phase space for resonance is desirable yet it is beyond the scope of the current study. We conclude that the application of the smoothed inflaton background \eqref{eq:theta dot approximation} to evaluate the Higgs relaxation \eqref{eq:eom_h_finiteT} is a good approximation for an inflaton mass $m_\phi \ll \Lambda_{I}^2/\Lambda$.

\section{Instability of Higgs mode functions}\label{AppendixB}
It is well-known that parametric resonance of perturbations in a field can be triggered via couplings with the oscillating inflaton or the non-harmonic oscillations of its VEV (zero-mode) motion, leading to unstable growth in some ranges of the mode functions. In this section, we examine if Higgs mode functions $\delta h_k$ exhibit resonance instability during forced relaxation. The equation of motion for the non-zero modes of the Higgs perturbations reads
\begin{align}\label{eq:eom_dh}
\ddot{\delta h_k}+3H\dot{\delta h_k}+\left( \frac{k^2}{a^2} + 3\lambda h^{2} - \frac{\rho_{\phi}^{}}{\Lambda^{2}} + \alpha_{T}^{2}T^{2}\right)\delta h_k=0,
\end{align}
where the zero-mode function $h$ and the inflaton density $\rho_\phi = \dot{\phi}^2$ are both oscillating during forced relaxation. 
For a general periodic function given by the combination of $h$ and $\rho_{\phi}$ in \eqref{eq:eom_dh}, the instability can be described by the Hill's equation (for example, see \cite{Fukunaga:2019unq}).

\begin{figure}
	\begin{center}
		\includegraphics[width=76mm]{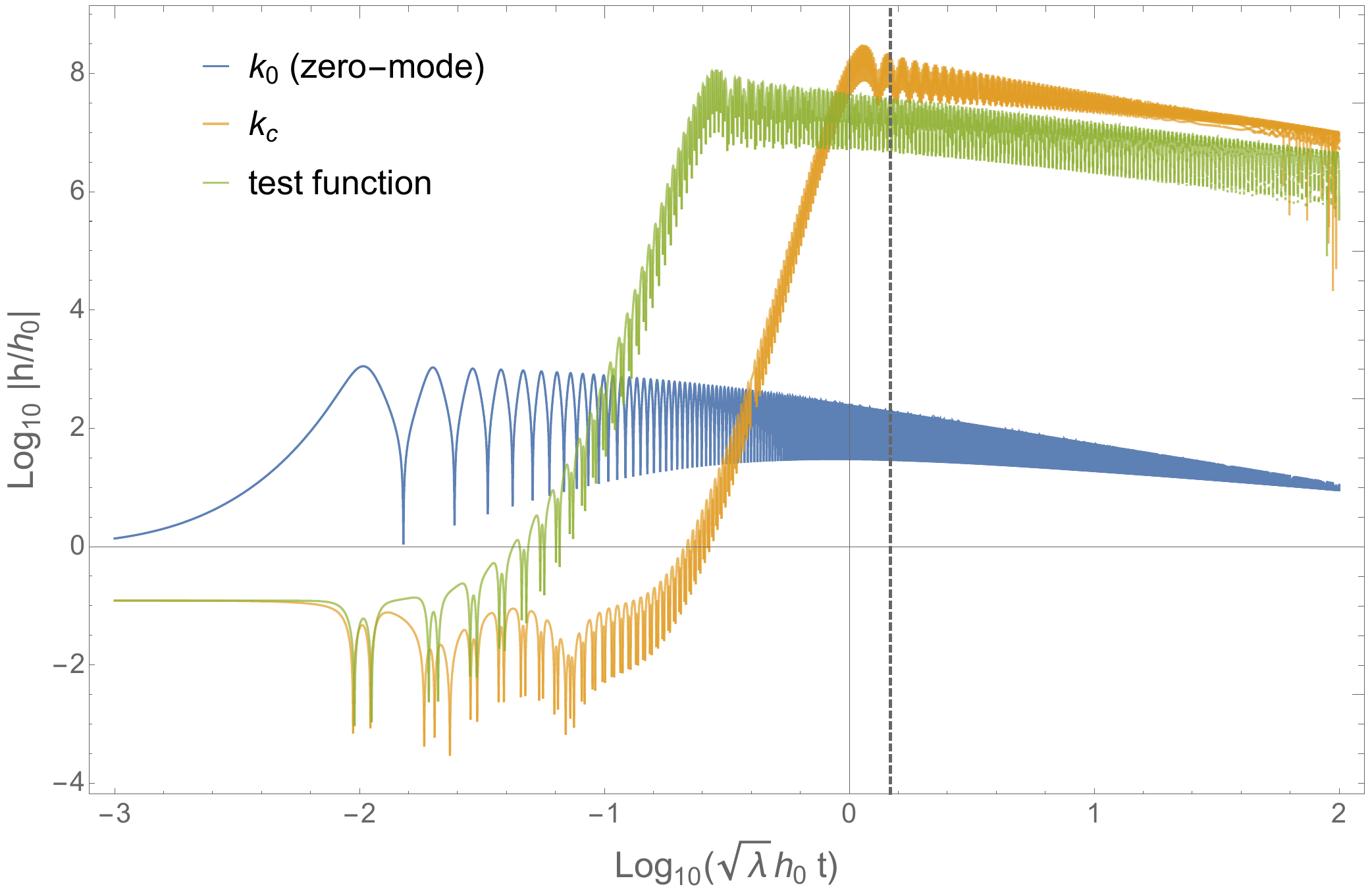}
		\hfill
		\includegraphics[width=76mm]{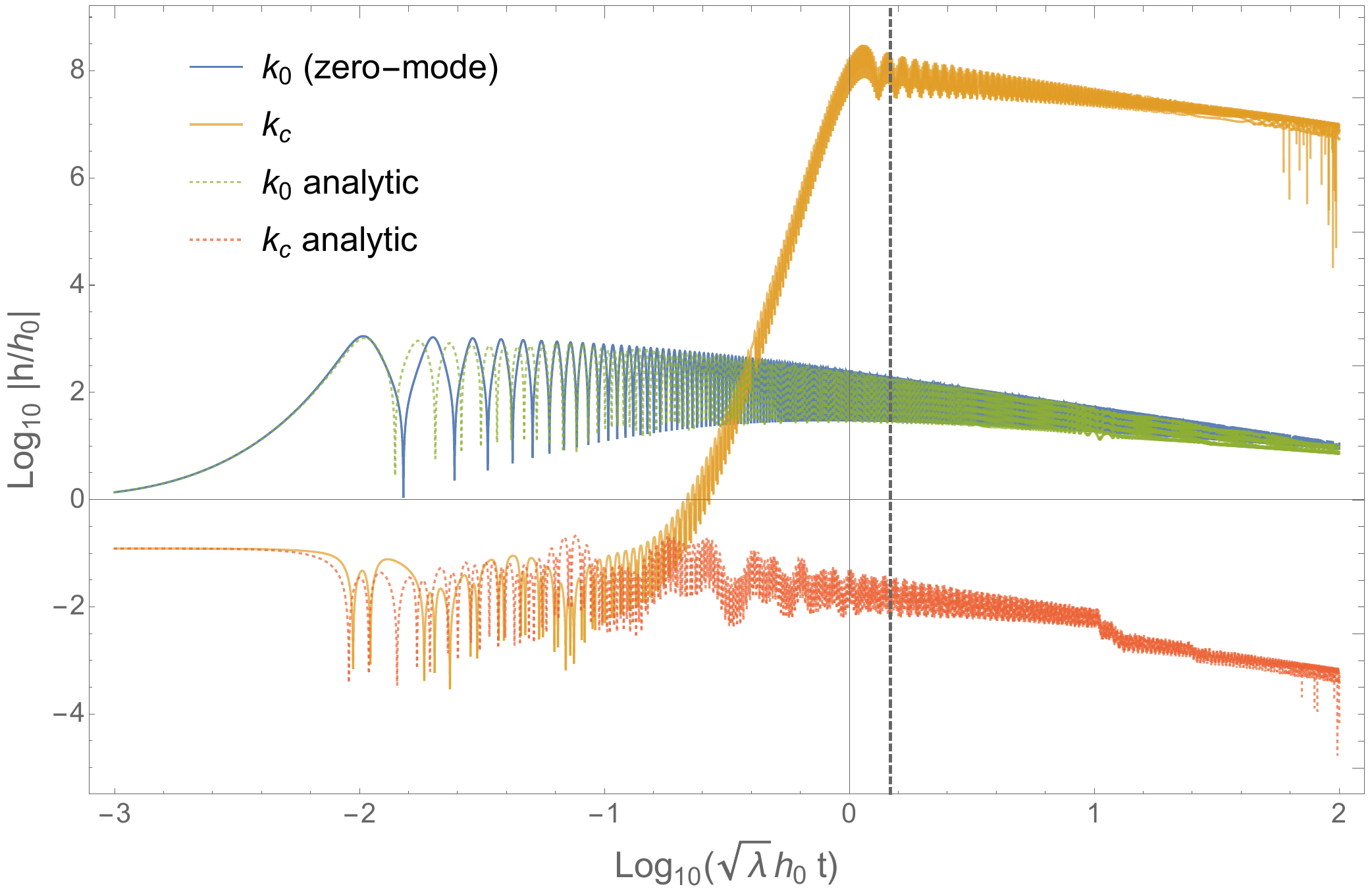}
	\end{center}
	\caption{\label{fig_resonance_dh}The evolution of Higgs mode functions with $\Lambda_{I} = 2.5\times 10^{15}$ GeV and $\Lambda = 2\times 10^{16}$ GeV. The results in the left panel are based on smoothed inflaton density \eqref{inflaton_density_smoothed}. The dotted lines in the right panel are solved by analytic inflaton density \eqref{eq:inflaton_analytic} with the mass parameter $m_\phi = 0.05 \Lambda_{I}^2/\Lambda$. The dashed line is the time when maximum temperature is reached.  
	}
\end{figure}

If we consider an inflaton mass $m_\phi \ll \Lambda_{I}^2/\Lambda$, the oscillating period of $\phi$ is much larger than $h$, according to the findings in 
Appendix~\ref{Appendix}. In this case one can observe that the most unstable mode arises around the critical value $k_c = \Lambda_{I}^2/\Lambda$, where $k_c^2/a^2 = \rho_{\phi}/\Lambda^2$ at $t = t_0 = 0$. To see this, let us first solve the evolution of the test mode function $\delta h_{\rm test}$ from
\begin{align}
\ddot{\delta h}_{\rm test}+3H\dot{\delta h}_{\rm test}+\left(  3\lambda h^{2} + \alpha_{T}^{2}T^{2}\right)\delta h_{\rm test} = 0.
\end{align} 
This equation coincides with the equation of motion for $k_c$ at the moment $t = 0$. The results in the left panel of Figure~\ref{fig_resonance_dh} indicate that $\delta h_{\rm test}$ is unstable as it exceeds the zero-mode $h$ before the maximal reheating temperature is reached. Here $h$ is solved by the smoothed denstiy $\rho_{\phi} = \Lambda_{I}^4a^{-3}e^{-\Gamma_{I}t}$, which corresponds to $m_\phi\rightarrow 0$. Away from $t = 0$, $k_c^2/a^2$ does not cancel exactly with $\rho_{\phi}/\Lambda^2$ in \eqref{eq:eom_dh} so that $\delta h_{k_c}$ grows slower than $\delta h_{\rm test}$. 

Recalling that the inflaton oscillation is nearly harmonic, we shall replace in \eqref{eq:eom_dh} the inflaton density by $\rho_{\phi}^{\rm analytic}$ and $h$ by the solution of \eqref{eq:eom_h_analytic} to obtain a more realistic evolution for $\delta h_k$. The results in the right panel of Figure~\ref{fig_resonance_dh} show that there is no resonance instability for the mode function around $k = k_c$ if the inflaton oscillation is taken into account.
This result implies that $\Lambda_{I}^2/\Lambda > m_\phi \gg H_I$ is a suitable mass range for forced relaxation.
We note that our initial value for $\delta h_k$ is estimated by the field condensation in a symmetry breaking potential based on the stochastic approach \cite{Starobinsky:1994bd}.

\section{Non-perturbative Higgs decay}\label{AppendixC}

\begin{figure}
	\begin{center}
		\includegraphics[width=76mm]{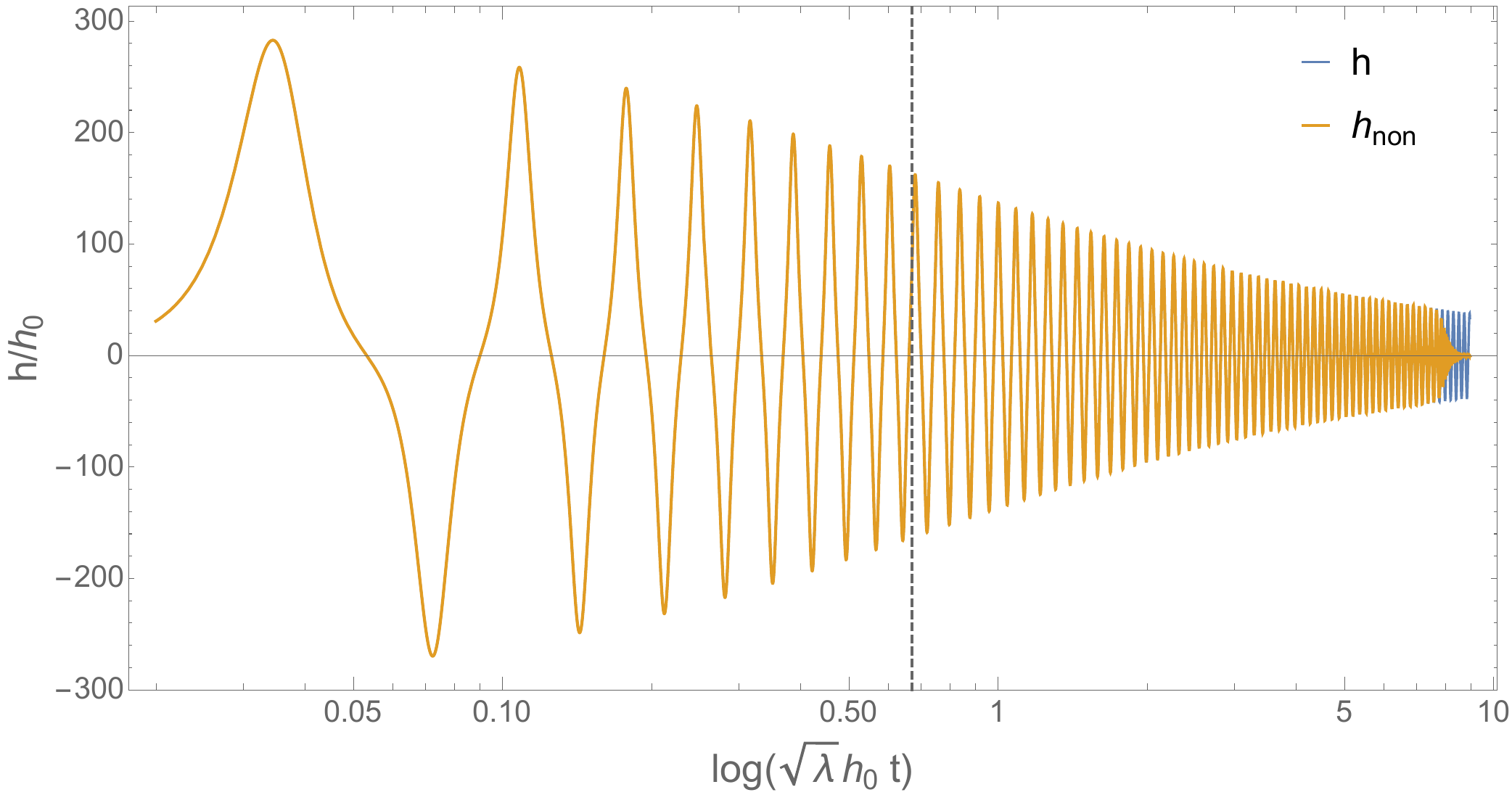}
		\hfill
		\includegraphics[width=76mm]{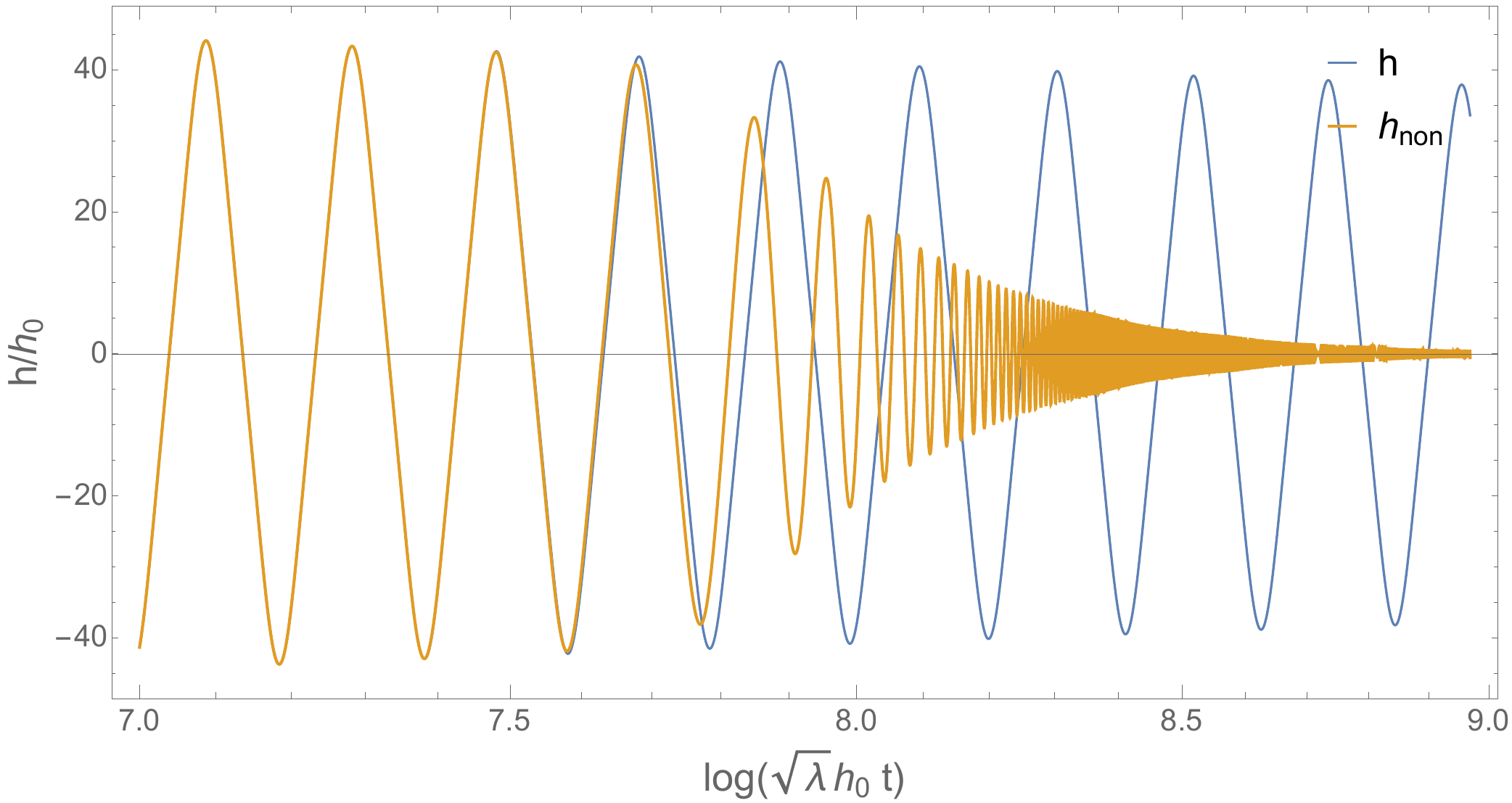}
	\end{center}
	\caption{\label{fig_non_p_decay}The evolution of Higgs mode functions with $\Lambda_{I} = 5\times 10^{15}$ GeV and $\Lambda = 2\times 10^{16}$ GeV. The blue (orange) line shows the evolution without (with) non-pertubative decay. The dashed line is the time when maximum temperature is reached.  
	}
\end{figure}

The oscillating Higgs VEV generates a time-dependent mass term for SM particles. Such a mass term as a periodic function in time can lead to non-perturbative production of particle modes that experience the parametric resonance. For scenarios of the free-fall relaxation \cite{Yang:2015ida}, the non-pertubative production of SM particles has negligible backreaction on the dynamics of the Higgs VEV before the maximal temperature is reached, and thus the non-pertubative Higgs decay has little impact on the final lepton asymmetry. In this section, we extend the study to the forced relaxation scenarios.

As a crude approximation for the non-perturbative effect, we consider the effective Higgs potential at tree-level as
\begin{align}
V_{\rm eff}(h) = \frac{1}{4}\lambda h^4 - \frac{1}{2} \frac{\rho_\phi}{\Lambda^2} h^2 + \frac{1}{2}\alpha_{T}^2 T^2 h^2,
\end{align}
where $\rho_{\phi}$ is the smoothed inflaton density \eqref{eq:theta dot approximation} valid for an inflaton mass $m_\phi \ll \Lambda_{I}^2/\Lambda$ (see Appendix~\ref{Appendix}). We focus on the dominant channels of Higgs decay into $W$ and $Z$ bosons, whose tree-level masses are led by
\begin{align}\label{gauge boson masses}
m_W^2 = g^2h^2/4, \qquad m_Z^2 = (g^2 + g^{\prime 2})h^2/4.
\end{align}
The running of couplings is not included in this approximation. The equation of motion for weak gauge fields is governed by the Lagrangian
\begin{align}
\mathcal{L}_A = \frac{1}{4}g^{\mu\nu}\left[g^2 W_\mu^+ W_\nu^- + \frac{1}{2}(g^2 +g^{\prime 2})Z_\mu Z_\nu\right] h^2 + \mathcal{L}_{A, \mathrm{kin}},
\end{align}
where the kinetic terms can be written as
\begin{align}
\mathcal{L}_{A, \mathrm{kin}} = -\frac{1}{2}(\nabla_\mu W^+_\nu - \nabla_\nu W_\mu^+)(\nabla^\mu W^{-\nu} - \nabla^\nu W^{-\mu}) -\frac{1}{4}(\nabla_\mu Z_\nu - \nabla_\nu Z_\mu)^2.
\end{align}
Here we ignore the non-linear corrections come from non-Abelian contributions \cite{Yang:2015ida,Enqvist:2013kaa,Enqvist:2014tta}.

We shall focus on the transverse mode $\vec{A}_T(\vec{k}, t)$ of the vector bosons $A_\mu = W_\mu^{\pm}$ or $Z_\mu$, since the longitudinal mode $\vec{A}_L(\vec{k}, t)$ (including $A_0$) is suppressed by an additional friction term due to the time-dependence of the masses $m_A = m_A(t)$ given in \eqref{gauge boson masses} \cite{Yang:2015ida,Enqvist:2014tta}. Note that $\vec{k}\cdot \vec{A}_T = 0$ and $\vec{k}\times \vec{A}_L = 0$. In terms of the conformal time $d\eta = a^{-1}dt$, the equation of motion for the transverse mode is
\begin{align}
\frac{d^2}{d\eta^2} \vec{A}_T + (k^2 + a^2 m_A^2) \vec{A}_T =0,
\end{align}
where $a(\eta)$ is given by the smoothed background $\rho_{\phi}$.

The number density of particles with momentum $\vec{k}$ is defined in the conventional manner
\begin{align}
n_k(\eta) \equiv \frac{1}{2 \omega_k} \left[\vert A_\mu^\prime(\vec{k},\eta)\vert^2 + \omega_k^2 \vert A_\mu(\vec{k}, \eta)\vert^2 \right] - \frac{1}{2},
\end{align}
where $\omega_k = (k^2 + a^2m_A^2)^{1/2}$ is the energy of each particle and $A^\prime \equiv \partial A/\partial\eta$. The initial conditions $A_T(k, 0) = 1/\sqrt{2\omega_k}$ and $A_T^\prime(k, 0) = -i \sqrt{\omega_k/2}$ satisfy $n_k(0) = 0$ and the Wronskian condition $AA^{\prime\ast} - A^\ast A^\prime = i$.

The backreaction of the non-perturbative decay to the Higgs dynamics is accounted for by the induced masses from the particle production as
\begin{align}\label{backreaction masses}
m_{h, W}^2 = -\frac{1}{2}g^2\langle W_\mu^+ W^{-\mu}\rangle, \quad m_{h, Z}^2 = -\frac{1}{4}(g^2 + g^{\prime 2}) \langle Z_\mu Z^\mu\rangle,
\end{align}
where the expectation value is approximated according to \cite{Yang:2015ida}:
\begin{align}
g^{\mu\nu} \langle A_\mu A_\nu \rangle \simeq \frac{-2}{a^2} \langle A_T^2\rangle \approx \frac{-1}{\pi^2 a^2} \int_{0}^{k_{\rm max}} \frac{n_k}{\omega_k} k^2dk.
\end{align}
The cutoff $k_{\rm max} \simeq a m_A/2 $ is estimated by assuming the existence of the first instability band in the Mathieu equation around the time of the first oscillatory period.

Our numerical check in Figure~\ref{fig_non_p_decay} shows that the backreaction due to non-perturbative Higgs decay is not important until the number density of produced particles is large enough so that the induced masses \eqref{backreaction masses} dominate the Higgs potential. We therefore conclude that the non-perturbative decay of the Higgs condensate has negligible effect on forced relaxation by the maximum-temperature time where most of the lepton asymmetry has been generated.

Before closing the section, we remark on the perturbative Higgs decay into SM fermions as considered in  \cite{Yang:2015ida,Enqvist:2013kaa}. One can easily check that during forced relaxation with $h/h_0 \gg 1$, the top quark mass $m_t = y_t h/\sqrt{2} \gg m_h$, where $y_t = \mathcal{O}(1)$ denote the top Yukawa coupling and $m_h^2 \equiv \partial^2 V_{\rm eff}(h)/\partial h^2 $. This means that the perturbative decay of Higgs into top quarks (the leading channel for free-fall relaxation) is kinematically blocked \cite{Enqvist:2013kaa}. For the leading channel of perturbative Higgs decay, we have numerically confirmed that the decay rate into bottom quarks $\Gamma(h \rightarrow b\bar{b}) \ll H$. This relation holds very well by the time when the temperature reaches the maximum. Here $\Gamma$ is estimated in Ref.~\cite{Yang:2015ida}. As a result, the perturbative decay (or thermalization) of the Higgs is not efficient during the epoch when the lepton asymmetry is generated.

\end{document}